\documentclass[preprint,12pt]{elsarticle}

\usepackage{amssymb}
\usepackage{amsmath}

\usepackage{subcaption}
\usepackage{booktabs}
\usepackage{tikz}

\usepackage{hyperref}

\usepackage{tikz}
\usetikzlibrary{calc,fit,positioning,shapes,decorations}
\usetikzlibrary{decorations.pathreplacing}
\usetikzlibrary{intersections}
\usetikzlibrary{tikzmark,decorations.pathreplacing,calligraphy}
\usepackage{pgfplots}
\usetikzlibrary{pgfplots.fillbetween}

\definecolor{col1}{HTML}{000000} 
\definecolor{col2}{HTML}{006DB2} 
\definecolor{col3}{HTML}{BD0006} 
\definecolor{col4}{HTML}{6BAD00} 
\definecolor{col5}{HTML}{FFC300} 

\tikzstyle{subdued}= [color=gray,very thin]
\tikzstyle{slantStyle}= [scale=2/3,xslant=1,yscale=0.5] 
\tikzset{
	>=latex	
}

\tikzset{
  invisible/.style={opacity=0},
  visible on/.style={alt={#1{}{invisible}}},
  alt/.code args={<#1>#2#3}{%
    \alt<#1>{\pgfkeysalso{#2}}{\pgfkeysalso{#3}} 
  },
}

\tikzstyle{uml}=[node distance=1em,draw,rounded corners=false,text centered,text width=7em, minimum width=7em, minimum height=4em,
inner xsep=3em,inner ysep=1em,align=center]
\tikzstyle{umlStart}=[uml,ellipse, inner sep=-1em]
\tikzstyle{umlEnd}=[uml,rounded corners=1.5em]
\tikzstyle{umlCommand}=[uml]
\tikzstyle{umlDecision}=[uml,diamond, aspect=2, align=center,inner xsep=1em, inner ysep=-1em,minimum width=12em]
\tikzstyle{umlRoutine}=[uml,rectangle split, rectangle split horizontal, rectangle split parts=3]
\tikzstyle{umlLoopStart}=[uml,trapezium, trapezium angle=60.0, inner xsep=0.8em]
\tikzstyle{umlLoopEnd}=[umlLoopStart, trapezium angle=-60.0]

\usetikzlibrary{patterns}
\usetikzlibrary{arrows.meta} 

\colorlet{xcol}{blue!70!black}
\colorlet{darkblue}{blue!40!black}
\colorlet{myred}{red!65!black}
\tikzstyle{mydashed}=[xcol,dashed,line width=0.25,dash pattern=on 2.2pt off 2.2pt]
\tikzstyle{axis}=[->,thick] 
\tikzstyle{ell}=[{Latex[length=3.3,width=2.2]}-{Latex[length=3.3,width=2.2]},line width=0.3]
\tikzstyle{dx}=[-{Latex[length=3.3,width=2.2]},darkblue,line width=0.3]
\tikzstyle{ground}=[preaction={fill,top color=black!10,bottom color=black!5,shading angle=20},
                    fill,pattern=north east lines,draw=none,minimum width=0.3,minimum height=0.6]
\tikzstyle{mass}=[line width=0.6,red!30!black,fill=red!40!black!10,rounded corners=1,
                  top color=red!40!black!20,bottom color=red!40!black!10,shading angle=20]
\tikzstyle{spring}=[line width=0.8,blue!7!black!80,snake=coil,segment amplitude=5,segment length=5,line cap=round]
\tikzset{>=latex} 
\tikzstyle{force}=[->,myred,very thick,line cap=round]

\usepackage[export]{adjustbox}

\newcommand{\figref}[1]{Fig.~\ref{#1}}
\usepackage{letltxmacro}
\LetLtxMacro{\originaleqref}{\eqref}
\renewcommand{\eqref}{Eq.~\originaleqref}

\newcommand{\mv}[1]{\vec{#1}}
\newcommand{\mm}[1]{\boldsymbol{#1}}

\newcommand{\del}[2]{\frac{\partial #1}{\partial #2}}

\newcommand{\heaviside}{\mathcal{H}}

\newcommand{\dt}{\Delta t}
\newcommand{\dx}{\Delta x}

\newcommand{\xt}{\mv{x}, t}
\newcommand{\fpp}{f_i(\mv{x} + \mv{c}_i \Delta t, t + \Delta t)}
\newcommand{\fpc}{f_i^c(\xt)}

\journal{Journal}

\begin{document}
\begin{frontmatter}



\title{Coupled Level-Set Lattice Boltzmann Method on Adaptive Cartesian Grids} 

\author[label1]{Julian Vorspohl}
\affiliation[label1]{organization={Institute of Aerodynamics and Chair of Fluid Dynamics},
                     addressline={{Wüllnerstra{\ss}e 5a}},
                     city={Aachen},
                     country={Germany}}
\affiliation[label2]{organization={J\"ulich Aachen Research Center for Simulation and Data Science, RWTH Aachen University and Forschungszentrum J\"ulich, J\"ulich, Germany},
                     city={J\"ulich},
                     country={Germany}}

\author[label1]{Yuxing Peng}
\author[label1,label2]{Matthias Meinke}
\author[label1]{Dominik Krug}
\author[label1,label2]{Wolfgang Schröder}

\begin{abstract}
A novel coupled level-set lattice Boltzmann method on adaptive Cartesian grids for simulating liquid-gas multiphase flows is presented. The approach addresses the inherent challenges of accurately modeling multiphase systems characterized by sharp interfaces and large density ratios. By employing separate solution algorithms for each fluid phase which are coupled through boundary conditions at the interface the method is more accurate and more efficient. The study highlights the advantages of using lattice Boltzmann methods together with level-set techniques to track interfaces effectively while facilitating adaptive mesh refinement. Applications to various test cases, e.g., immiscible stratified flow and rising bubbles, demonstrate the method's capability to capture complex interfacial dynamics and validate its accuracy against literature data.
\end{abstract}


\end{frontmatter}


\section{Introduction}

Multiphase flows, which are characterized by the dynamic interaction of two or more immiscible fluid phases such as liquid and gas, are ubiquitous in nature and industrial applications, ranging from geysers and bubbly flows in chemical reactors to gas-leakage flows in oil-filled sealing gaps of positive displacement compressors~\cite{Legendre2025,Hurwitz2017,Nikolov2023}. Accurate numerical simulation of such complex flows is crucial for understanding, optimization, and predictive design. Since such flows are characterized by sharp interfaces, large density ratios, and intricate topological changes characteristic of such flows. It is still quite a challenge to satisfy those conditions.

At a fundamental level, approaches to simulate multiphase flows generally fall into two broad categories: one-fluid methods and two-fluid methods. Two-fluid models~\cite{Lauwers2021} typically treat each phase as an interpenetrating continuum, using separate sets of governing equations for each phase that are coupled through interfacial terms. While computationally efficient for dispersed flows, they often require closure relations for the interfacial momentum and mass exchange. For interface-resolved simulations, virtually no two-fluid approaches are discussed in the literature.

\paragraph{One-fluid methods}

In contrast, one-fluid methods~\cite{Sitompul2019, Fakhari2016, Chen2014, Shao2014} employ a single set of momentum equations across the entire domain. An indicator function like volume fraction or phase field is used to distinguish the phases and locate the interface. This approach intrinsically satisfies momentum conservation and is well-suited for simulating complex, deforming interfaces. However, this approach suffers from stability problems for higher density and viscosity ratios due to the steep gradients at the phase boundary. This requires special measures like filtering the velocity field or solving the pressure evolution equation.

A critical aspect of one-fluid simulations is the interface tracking technique. Methods like Volume-of-fluid (VOF) track the interface implicitly by advecting the volume fraction function, often requiring sophisticated reconstruction techniques, e.g., piecewise linear interface colculation (PLIC) to maintain interface sharpness~\cite{Hirt1981}. One of the main challenges for VOF methods is the curvature determination, due to the sharp nature of the solution, which requires additional algorithms to obtain smooth values~\cite{Han2024}. Level-set (LS) methods define the interface as the zero level of a smooth signed distance function, which simplifies the calculation of geometric properties like curvature but necessitates reinitialization to maintain the distance function property~\cite{Chen1997}. More recently, phase field (PF) methods, based on Cahn-Hilliard or similar equations, have emerged~\cite{Hu2020}, where the interface is treated as a thin, smooth transition zone dictated by a thermodynamic free energy functional. PF methods naturally handle topological changes like merging and breakup without explicit tracking.

While traditionally these approaches have been employed in connection with finite volume or finite difference solvers, novel approaches using the lattice Boltzmann method (LBM)~\cite{Kruger2017} became more popular as a numerical approach for fluid dynamics including multiphase flow. LBM simulates fluid behavior by tracking the evolution of mesoscopic particle distribution functions on a discrete lattice, offering computational advantages in parallelization and handling complex boundaries. For liquid-gas systems, LBM offers several distinct multiphase models, most of them fall into the one-fluid category. The pseudo-potential model or 'Shan-Chen' type model~\cite{Chen2014} is one of the most widely used LBM multiphase models. It introduces non-local forcing terms, the so-called `pseudo-potential', into the evolution equation to model inter-particle interactions, leading to phase separation and generating surface tension. Its simplicity and natural handling of large density ratios make it popular, though it often suffers from spurious currents near the interface.

The free-energy model or 'Swift' type model~\cite{Shao2014} is directly derived from the free-energy functional of fluid mixtures, e.g., Cahn-Hilliard or van der Waals equation of state, which is more thermodynamically consistent, allowing for the simulation of systems close to equilibrium and phase transitions. It generally offers greater accuracy in determining surface tension and bulk properties, but is often more computationally demanding.

Another class of one-fluid LB methods for multiphase flows with large density ratios is developed around the work of He et al. \cite{He1998}, who introduced a modified version of the discrete Boltzmann equation to account for non-ideal fluids. Two notable multiphase methods building on that idea are the work of Fakhari et al. \cite{Fakhari2016} and Sitompul et al. \cite{Sitompul2019}. In the first work \cite{Fakhari2016}, an LBM based phase field method is utilized to track the interface. For the mass and momentum equation, the modified discrete modified Boltzmann equation is extended further to account for the gradients in the density field and the surface tension, which is modelled by the continuous surface force (CSF) model \cite{Brackbill1992}. The resulting LBM method features multiple additional terms which have to be computed in the \textit{entire} computational domain, which involves high order gradients of the density and phase field using finite difference schemes. This largely negates the strong locality that normally makes LBM computationally efficient.
In the second work \cite{Sitompul2019}, an extension of the vanilla PF method is employed to track multiple bubbles individually. The discrete Boltzmann equation is modified accordingly to decouple pressure and density. The pressure for each time step is obtained by iteratively solving a Poission equation and subsequently applying a spatial filter to stabilize the solution. The lattice Boltzmann methods presented in the literature suitable for high density and viscosity ratios rely on modified versions of the discrete Boltzmann equation, which in general decreases the computational efficiency since the locality of the algorithm is weakened in the entire domain.\\

In summary, all one-fluid approaches in literature share this inherent drawback, as they all use \textit{one} solution algorithm for the entire domain. This leads to potentially suboptimal performance since the algorithm can be tailored to the requirements of one fluid phase only. Additional, non-local regularization terms are needed for the simulation of large density and viscosity ratios to stabilize the solution. In this manuscript, we introduce a two-fluid method on adaptive Cartesian grids for the simulation of liquid-gas multiphase flow that addresses these limitations by providing a general framework in which different standard solution algorithms can be employed for each fluid phase, which are coupled at a sharp interface through boundary conditions.\\

The manuscript is structured as follows. The novel numerical approach as well as the underlying data structures are introduced in Sec.\ref{sec:method}. The simulation setups and the discussion of the results are given in Sec.\ref{sec:results}. Finally, the findings are summarized and an outlook is presented in Sec.\ref{sec:summary}.

\section{Numerical Approach}
\label{sec:method}
A fully coupled two-fluid approach for the simulation of liquid-gas two-phase flows is employed. The flow field in both phases is simulated using a separate, standard solution algorithm, i.e., no non-local regularization terms to stabilize the solution at the phase interface are necessary. Considering the different physical and numerical properties of each phase, each solution algorithm is formulated such that it satisfies the accuracy and efficiency requirements of its relevant phase. This approach enables the independent choice of solution schemes. In the following, this is shown by using different lattice Boltzmann (LB) methods for the gas and liquid phases. Note that this approach is not limited to LB methods and could be extended to, e.g., finite volume (FV) methods for the gas phase, yielding a proper compressible solution. Here, the level-set based method described in~\cite{Schneiders2016} could be used. The interface between both phases is captured using a level-set method. Continuity, the effect of surface tension, and the jump condition in the stress tensor across the interface are enforced by a modified boundary condition for the liquid and gas flow solvers. All solution algorithms operate on a joint hierarchical Cartesian grid, which enables the efficient exchange of coupling data between all solvers without any I/O needed, and adaptive mesh refinement. In the following, all numerical methods as well as their coupling are briefly discussed.\\

The simulations including grid generation and post-processing are performed using m-AIA, an open source solver framework with a focus on fluid mechanics~\cite{Code}.

\subsection{Lattice Boltzmann method}
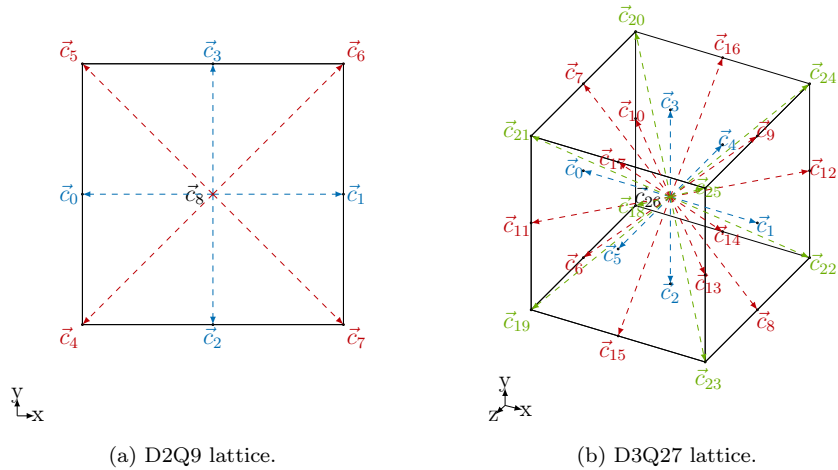
\begin{figure}[ht!]
    \centering
    \begin{subfigure}[t]{0.45\textwidth}
    \centering
    \scalebox{0.75}{\begin{tikzpicture}[scale=0.25*\textwidth/1cm]
  \newcommand\DX{1.5}
  \newcommand\DY{1.5}
  \newcommand\DZ{-0.0}
  \newcommand{\cor}[4]{\coordinate (#1)  at (#2*\DX, #3*\DY);}

	\cor{0}	{-1}{ 0}{ 0}
	\cor{1}	{ 1}{ 0}{ 0}
	\cor{2}	{ 0}{-1}{ 0}
	\cor{3}	{ 0}{ 1}{ 0}

	\cor{4}	{-1}{-1}{ 0}
	\cor{5}	{-1}{ 1}{ 0}
	\cor{6}	{ 1}{1}{ 0}
    \cor{7}	{ 1}{-1}{ 0}

	\cor{8}	{ 0}{ 0}{ 0}

  \cor{c0}{-1}{-1}{ 1};
  \cor{c1}{ 1}{-1}{ 1};
  \cor{c2}{ 1}{ 1}{ 1};
  \cor{c3}{-1}{ 1}{ 1};
  \cor{c4}{-1}{-1}{-1};
  \cor{c5}{ 1}{-1}{-1};
  \cor{c6}{ 1}{ 1}{-1};
  \cor{c7}{-1}{ 1}{-1};

  \draw[very thin] (c0)--(c1)--(c2)--(c3)--cycle; 
  \draw[very thin] (c0)--(c1)--(c5)--(c4)--cycle; 
  \draw[very thin] (c0)--(c4)--(c7)--(c3)--cycle; 
  \draw[very thin] (c1)--(c5)--(c6)--(c2)--cycle; 
  \draw[very thin] (c3)--(c2)--(c6)--(c7)--cycle; 

  \draw[fill=red] (8) circle (0.01) node[left]{$\mv{c}_{8}$};
  \foreach \p in {0,1,2,3}{
    \draw[dashed,->, col2] (8) -- (\p) node[pos=1.1]{$\mv{c}_{\p}$};
    \draw[fill=col2] (\p) circle (0.01);
 	 }
  \foreach \p in {4,5,6,7}{
    \draw[dashed,->, col3] (8) -- (\p) node[pos=1.1]{$\mv{c}_{\p}$};
    \draw[fill=col3] (\p) circle (0.01);
 	 }

  \draw[->] (-1.5*\DX + +1.5*\DZ, -1.7*\DY + +1.5*\DZ) -- +(0.2,0) node[pos=1.2]{x};
  \draw[->] (-1.5*\DX + +1.5*\DZ, -1.7*\DY + +1.5*\DZ) -- +(0,0.2) node[pos=1.2]{y};


\end{tikzpicture}}
    \caption{D2Q9 lattice.}
    \end{subfigure}
    \begin{subfigure}[t]{0.45\textwidth}
    \centering
    \scalebox{0.75}{\begin{tikzpicture}[scale=0.25*\textwidth/1cm]
  \newcommand\DX{1}
  \newcommand\DY{1}
  \newcommand\DZ{-0.6}
  \newcommand{\cor}[4]{\coordinate (#1)  at (#2*\DX + #4*\DZ, #3*\DY + #4*\DZ - 0.3*#2);}

	\cor{0}	{-1}{ 0}{ 0}
	\cor{1}	{ 1}{ 0}{ 0}
	\cor{2}	{ 0}{-1}{ 0}
	\cor{3}	{ 0}{ 1}{ 0}
	\cor{4}	{ 0}{ 0}{-1}
	\cor{5}	{ 0}{ 0}{ 1}

	\cor{6}	{-1}{-1}{ 0}
	\cor{7}	{-1}{ 1}{ 0}
	\cor{8}	{ 1}{-1}{ 0}
	\cor{9}	{ 1}{ 1}{ 0}

	\cor{10}	{-1}{ 0}{-1}
	\cor{11}	{-1}{ 0}{ 1}
	\cor{12}	{ 1}{ 0}{-1}
	\cor{13}	{ 1}{ 0}{ 1}
	
	\cor{14}	{ 0}{-1}{-1}
	\cor{15}	{ 0}{-1}{ 1}
	\cor{16}	{ 0}{ 1}{-1}
	\cor{17}	{ 0}{ 1}{ 1}

	\cor{18}	{-1}{-1}{-1}
	\cor{19}	{-1}{-1}{ 1}
	\cor{20}	{-1}{ 1}{-1}
	\cor{21}	{-1}{ 1}{ 1}
	\cor{22}	{ 1}{-1}{-1}
	\cor{23}	{ 1}{-1}{ 1}
	\cor{24}	{ 1}{ 1}{-1}
	\cor{25}	{ 1}{ 1}{ 1}

	\cor{26}	{ 0}{ 0}{ 0}

  \cor{c0}{-1}{-1}{ 1};
  \cor{c1}{ 1}{-1}{ 1};
  \cor{c2}{ 1}{ 1}{ 1};
  \cor{c3}{-1}{ 1}{ 1};
  \cor{c4}{-1}{-1}{-1};
  \cor{c5}{ 1}{-1}{-1};
  \cor{c6}{ 1}{ 1}{-1};
  \cor{c7}{-1}{ 1}{-1};

  \draw[very thin] (c0)--(c1)--(c2)--(c3)--cycle; 
  \draw[very thin] (c0)--(c1)--(c5)--(c4)--cycle; 
  \draw[very thin] (c0)--(c4)--(c7)--(c3)--cycle; 
  \draw[very thin] (c1)--(c5)--(c6)--(c2)--cycle; 
  \draw[very thin] (c3)--(c2)--(c6)--(c7)--cycle; 

  \draw[fill=red] (26) circle (0.01) node[left]{$\mv{c}_{26}$};
  \foreach \p in {0,1,2,3,4,5}{
    \draw[dashed,->, col2] (26) -- (\p) node[pos=1.1]{$\mv{c}_{\p}$};
    \draw[fill=col2] (\p) circle (0.01);
 	 }
  \foreach \p in {6,7,8,9,10,11,12,13,14,15,16,17}{
    \draw[dashed,->, col3] (26) -- (\p) node[pos=1.1]{$\mv{c}_{\p}$};
    \draw[fill=col3] (\p) circle (0.01);
 	 }
  \foreach \p in {18,19,20,21,22,23,24,25}{
    \draw[dashed,->,col4] (26) -- (\p) node[pos=1.1]{$\mv{c}_{\p}$};
    \draw[fill=col4] (\p) circle (0.01);
 	 }

  \draw[->] (-1*\DX + +1.5*\DZ, -1.5*\DY + +1.5*\DZ) -- +(0.2,-0.05) node[pos=1.2]{x};
  \draw[->] (-1*\DX + +1.5*\DZ, -1.5*\DY + +1.5*\DZ) -- +(0,0.2) node[pos=1.2]{y};
  \draw[->] (-1*\DX + +1.5*\DZ, -1.5*\DY + +1.5*\DZ) -- +(-0.1-0.01,-0.1) node[pos=1.2]{z};


\end{tikzpicture}}
    \caption{D3Q27 lattice.}
    \end{subfigure}
    \caption{Velocity space discretization in two and three dimensions.}
    \label{fig:d3q27}
\end{figure}
Based on the Boltzmann equation from kinetic gas theory, the lattice Boltzmann (LB) method was derived by discretizing in velocity space. The equation was split into a "streaming" and a "collision" step. The general form reads
\begin{equation}
    \fpp = \fpc = f_i(\xt) + \Omega_i(\mv{f}) + F_{i},
\end{equation}
with $f$ being the particle probability density function (PPDF), $\mv{c}_i$ the discrete particle velocity for direction $i$, $f^c$ the post-collision state, $\Omega_i$ the collision operator, and $F_i$ the forcing term. In the following, the PPDF $f_i(\xt)$ is simply referred to as $f_i$. In two dimensions, a lattice consisting of 9 discrete velocities (D2Q9) is used, while the three-dimensional discretization is based on 27 discrete velocities (D3Q27) (\figref{fig:d3q27}). By calculating the moments of the PPDF, the density $\rho$ and velocity $\mv{u}$ are obtained in discrete form
\begin{equation}
    \rho = \sum_i f_i(\xt)
    \quad \text{and} \quad
    \mv{u} =  \frac{1}{\rho} \sum_i \mv{c}_i f_i(\xt).
    \label{eq:macro}
\end{equation}
For the collision operator, different options are available in the literature. The Bhatnagar-Gross-Krook (BGK) operator~\cite{BGK} and the cumulant collision operator \cite{Geier2015} are briefly discussed. The BGK collision operator, which is suitable for low Reynolds number flows, reads
\begin{equation}
    \Omega_i(\mv{f}) = \omega_{BGK} (f_i^{eq} - f_i),
\end{equation}
with the Maxwell equilibrium distribution function $f_i^{eq}$ given by
\begin{equation}
    f_i^{eq} = w_i \rho \left[1 + \frac{\mv{c}_i \cdot \mv{u}}{c_s^2} + \frac{(\mv{c}_i \cdot \mv{u})^2}{2c_s^4} - \frac{\mv{u} \cdot \mv{u}}{2c_s^2}\right],
    \label{eq:feq}
\end{equation}
where $w_i$ are the weighting factors. In the two-dimensional case, these are 4/9, 1/9, and 1/36 for the center, the Cartesian, and the edge diagonal. For three dimensions, the weighting factors are 8/27, 2/27, 1/54, and 1/216 for the center, the Cartesian, the edge diagonal, and space diagonal. The speed of sound is denoted as $c_s$. The collision frequency $\omega_{BGK}$ is given by
\begin{equation}
    \omega_{BGK} = \frac{\dt c_s^2}{\nu + \frac12 \dt c_s^2},
    \label{eq:omega}
\end{equation}
with $\nu$ denoting the viscosity of the fluid. The forcing term is
\begin{equation}
    F_i = - \frac{\rho g}{c_s^2} w_i \mv{c}_i \cdot \mv{e}_z
\end{equation}
to account for the effect of gravity $g$.\\

For high Reynolds number flows in three dimensions, the cumulant collision operator has been developed in~\cite{Geier2015}. The PPDF is transformed into countable cumulants $c_\alpha$ to obtain Galilean invariant and statistically independent quantities. The collision is then performed for the cumulants, which reads
\begin{equation}
    c_\alpha^c = c_\alpha + \omega_\alpha (c^{eq}_\alpha - c_\alpha),
\end{equation}
where $c_\alpha^{eq}$ denotes the Maxwell equilibrium distribution~\eqref{eq:feq} in cumulant space and $c_\alpha^c$ the post-collision cumulants. The relaxation parameters in cumulant space are
\begin{equation}
    \omega_1 = \omega_{BGK},\quad \omega_\alpha = 1\quad \text{for}\quad \alpha \neq 1.
\end{equation}
For solid walls, the interpolated bounce back method ~\cite{Bouzidi2001} is used. The missing PPDFs from the boundary are constructed as
\begin{equation}
    f_{\Bar{i}}(\xt+1) = 
    \begin{cases}
        2q_if_i^c(\xt) + (1-2q_i)f_i^c(\mv{x}-\mv{c}_i, t) + \Delta f_i&\text{if $q_i<\frac12$}\\
        \frac{1}{2q_i}f_i^c(\xt) + \frac{2q_i-1}{2q_i}f_{\Bar{i}}^c(\mv{x}-\mv{c}_i, t) + \frac{1}{2q_i} \Delta f_i&\text{if $q_i\geq \frac12$,}
        \label{eq:bb}
    \end{cases}
\end{equation}
where $\Bar{i}$ denotes the opposite direction of $i$ and $q_i$ the normalized distance along $\mv{e}_i$ from the cell center to the wall. The post-collision state is denoted by $f_i^c(\mv{x}-\mv{c}_i, t)$. For static walls, the momentum source term $\Delta f_i$ is zero, for walls moving with the velocity $\mv{u}_B$ it is given by
\begin{equation}
    \Delta f_i = \frac{2 w_i}{c_s^2} (\mv{c}_i \cdot \mv{u}_B).
\end{equation}
To support local grid refinement in the lattice Boltzmann method, the approach of Dupuis and Chopard is employed~\cite{Dupuis2003}. Missing distributions at the refinement jump are obtained by spatial and temporal interpolation between the known neighboring distributions. Since $\dt \approx \dx$, the collision frequency for different grid spacings is adapted to ensure that the viscosity $\nu$ is constant~\eqref{eq:omega}.
\subsection{Level-set method}
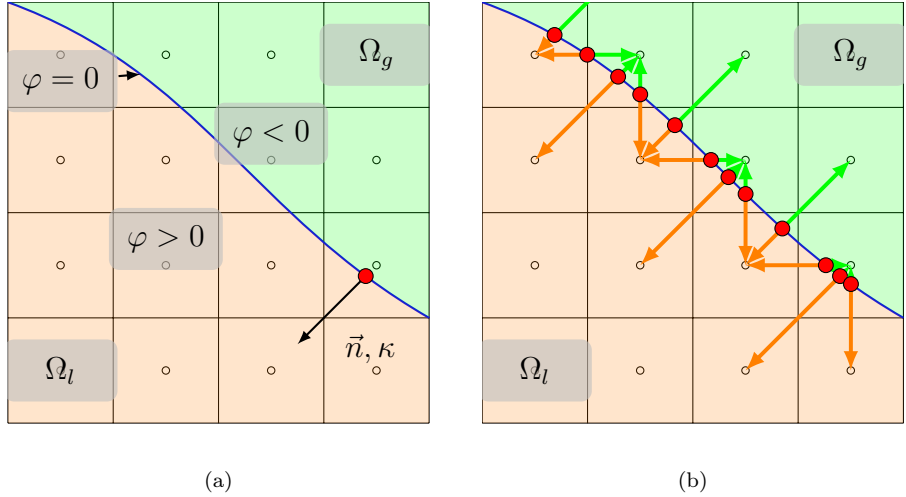
\begin{figure}
    \centering
    \begin{subfigure}[t]{0.45\textwidth}
    \centering
    \begin{tikzpicture}[scale=1.4,every node/.style={minimum size=1cm},on grid, block/.style ={rectangle, thick, fill=black!25, opacity=0.6, text width=3em, align=center, rounded corners, minimum height=2em}]
   \begin{scope}
   	   \clip (0, 0) rectangle (4,4);
       \draw[step=1, black] (0,0) grid (4,4); 
       
	   \foreach \x in {0.5,1.5,2.5,3.5}
	   \foreach \y in {0.5,1.5,2.5,3.5}
       \draw (\x,\y) circle (1pt);       
       
       \draw [thick, blue, name path=bnd] (0,4) to[out=-20,in=150] (4,1);
       \path [thick, blue, name path=lower] (0,0) to (4,0);
       \tikzfillbetween[of=bnd and lower, on layer=]{orange, opacity=0.2};
       \path [thick, blue, name path=upper] (0,4) to (4,4);
       \tikzfillbetween[of=bnd and upper, on layer=]{green, opacity=0.2};
       
       \draw (0.5,0.5) node[block] (F1) {};
	   \node at (F1) {$\Omega_l$}; 
       \draw (3.5,3.5) node[block] (F2) {};
       \node at (F2) {$\Omega_g$};
       
       \draw (1.5,1.75) node[block] (phi1) {};
       \node at (phi1) {$\varphi>0$};
       \draw (2.5,2.75) node[block] (phi2) {};
       \node at (phi2) {$\varphi<0$};
       \draw (0.5,3.25) node[block] (phi0) {};
       \node at (phi0) {$\varphi=0$};
       \coordinate (A) at (3.5, 3.5);
       \path[name path=link] (phi0)--(A);
       \draw[->, thick, name intersections={of=link and bnd,by={B}}](phi0)--(B);
       
       \coordinate (A6) at (2.75, 0.75);
       \coordinate (B6) at (3.5, 1.5);
       \path[name path=link6] (A6)--(B6);
       \draw[->, thick, name intersections={of=link6 and bnd,by={C6}}](C6)-- node[midway,auto] {$\vec{n},\kappa$}(A6);  
       \draw[fill=red] (C6) circle (2pt);
   \end{scope}
%
%
%
\end{tikzpicture}
    \vspace{-1.5em}
    \caption{}
    \label{fig:coupling:ls}
    \end{subfigure}
    \begin{subfigure}[t]{0.45\textwidth}
    \centering
    \begin{tikzpicture}[scale=1.4,every node/.style={minimum size=1cm},on grid, block/.style ={rectangle, thick, fill=black!25, opacity=0.6, text width=3em, align=center, rounded corners, minimum height=2em}, disto/.style={->, orange, line width=1.5pt},
distg/.style={->, green, line width=1.5pt}]
   \begin{scope}
   	   \clip (0, 0) rectangle (4,4);
       \draw[step=1, black] (0,0) grid (4,4); 
       
	   \foreach \x in {0.5,1.5,2.5,3.5}
	   \foreach \y in {0.5,1.5,2.5,3.5}
       \draw (\x,\y) circle (1pt);       
       
       \draw [thick, blue, name path=bnd] (0,4) to[out=-20,in=150] (4,1);

	   \path [thick, blue, name path=lower] (0,0) to (4,0);
       \tikzfillbetween[of=bnd and lower, on layer=]{orange, opacity=0.2};
       \path [thick, blue, name path=upper] (0,4) to (4,4);
       \tikzfillbetween[of=bnd and upper, on layer=]{green, opacity=0.2};

       \coordinate (A1) at (0.5, 3.5);
       \coordinate (A2) at (0.5, 2.5);
       \coordinate (A3) at (1.5, 2.5);
       \coordinate (A4) at (1.5, 1.5);
       \coordinate (A5) at (2.5, 1.5);
       \coordinate (A6) at (2.5, 0.5);
       \coordinate (A7) at (3.5, 0.5);
      
       \coordinate (B1) at (1.5, 4.5);
       \coordinate (B2) at (1.5, 3.5);
       \coordinate (B3) at (2.5, 3.5);
       \coordinate (B4) at (2.5, 2.5);
       \coordinate (B5) at (3.5, 2.5);
       \coordinate (B6) at (3.5, 1.5);
      
       \path[name path=link11] (A1)--(B1);
       \draw[disto, name intersections={of=link11 and bnd,by={C11}}](C11)--(A1);
       \draw[distg](C11)--(B1);
       \draw[fill=red] (C11) circle (2pt);
       \path[name path=link12] (A1)--(B2);
       \draw[disto, name intersections={of=link12 and bnd,by={C12}}](C12)--(A1);
       \draw[distg](C12)--(B2);
       \draw[fill=red] (C12) circle (2pt);

       \path[name path=link22] (A2)--(B2);
       \draw[disto, name intersections={of=link22 and bnd,by={C22}}](C22)--(A2);
       \draw[distg](C22)--(B2);
       \draw[fill=red] (C22) circle (2pt);

       \path[name path=link33] (A3)--(B2);
       \draw[disto, name intersections={of=link33 and bnd,by={C32}}](C32)--(A3);
       \draw[distg](C32)--(B2);
       \draw[fill=red] (C32) circle (2pt);
       \path[name path=link33] (A3)--(B3);
       \draw[disto, name intersections={of=link33 and bnd,by={C33}}](C33)--(A3);
       \draw[distg](C33)--(B3);
       \draw[fill=red] (C33) circle (2pt);
       \path[name path=link34] (A3)--(B4);
       \draw[disto, name intersections={of=link34 and bnd,by={C34}}](C34)--(A3);
       \draw[distg](C34)--(B4);
       \draw[fill=red] (C34) circle (2pt);

       \path[name path=link44] (A4)--(B4);
       \draw[disto, name intersections={of=link44 and bnd,by={C44}}](C44)--(A4);
       \draw[distg](C44)--(B4);
       \draw[fill=red] (C44) circle (2pt);
       
       \path[name path=link54] (A5)--(B4);
       \draw[disto, name intersections={of=link54 and bnd,by={C54}}](C54)--(A5);
       \draw[distg](C54)--(B4);
       \draw[fill=red] (C54) circle (2pt);
       \path[name path=link55] (A5)--(B5);
       \draw[disto, name intersections={of=link55 and bnd,by={C55}}](C55)--(A5);
       \draw[distg](C55)--(B5);
       \draw[fill=red] (C55) circle (2pt);
       \path[name path=link56] (A5)--(B6);
       \draw[disto, name intersections={of=link56 and bnd,by={C56}}](C56)--(A5);
       \draw[distg](C56)--(B6);
       \draw[fill=red] (C56) circle (2pt);
       
       \path[name path=link6] (A6)--(B6);
       \draw[disto, name intersections={of=link6 and bnd,by={C6}}](C6)--(A6);
       \draw[distg](C6)--(B6);
       \draw[fill=red] (C6) circle (2pt);
       
       \path[name path=link76] (A7)--(B6);
       \draw[disto, name intersections={of=link76 and bnd,by={C76}}](C76)--(A7);
       \draw[distg](C76)--(B6);
       \draw[fill=red] (C76) circle (2pt);
	   
	   \draw (0.5,0.5) node[block] (F1) {};
	   \node at (F1) {$\Omega_l$}; 
       \draw (3.5,3.5) node[block] (F2) {};
       \node at (F2) {$\Omega_g$}; 
   \end{scope}
%
%
%
\end{tikzpicture}
    \vspace{-1.5em}
    \caption{}
    \label{fig:coupling:lb}
    \end{subfigure}
    \caption{Interface description using the level-set (a). Missing distributions for both lattice Boltzmann solvers ({\protect\tikz\protect\draw[->,orange,line width=1.0pt](0, 0.0) -- (0.3,0.3);}/{\protect\tikz\protect\draw[->,green,line width=1.0pt](0, 0.0) -- (0.3,0.3);})
    (b). Intersection between distributions and the interface are marked with {\protect\tikz\protect\draw[fill=red] (0, 0) circle (2pt);}.}
    \label{fig:coupling}
\end{figure}
To capture the interface between the liquid and gas fluid phases, a level-set method is applied \cite{Sethian1987}. The scalar variable $\varphi$ is introduced to identify the computational domain for each phase and the interface between them, such that
\begin{equation}
    \varphi(\xt)
    \begin{cases}
        < 0 \quad \Omega_{l} \quad \text{liquid phase}\\
        = 0 \quad \Gamma_{i} \quad \text{interface}\\
        > 0 \quad \Omega_{g} \quad \text{gas phase},
    \end{cases}
\end{equation}
as shown \figref{fig:coupling:ls}.
Additionally, the level-set is initialized and subsequently maintained such that $\varphi$ represents the signed distance to the interface. The temporal evolution of the interface is predicted by advecting the scalar field $\varphi$ with the local fluid velocity which yields
\begin{equation}
    \del{\varphi}{t} + \mv{\tilde{u}} \cdot \nabla \varphi = 0.
\end{equation}
To solve this transport equation, a fifth-order upwind-central scheme for the spatial discretization and a third-order total variation diminishing Runge-Kutta scheme for the time integration is employed~\cite{TVD}. Note that the transport equation is only solved in a narrow band around the interface~$\Gamma$, as described in~\cite{Sethian1987}. In general, the velocity field $\mv{\tilde{u}}$ used for the transport of the interface is only known at the interface itself. By using so-called hyperbolic extension, the velocity $\mv{u}_I$ is extended in the interface normal direction~\cite{Chen1997, Peng1999} to the entire level-set tube. This is done by solving an auxiliary first-order PDE in artificial time $\tau$ given by
\begin{equation}
    \frac{\partial \mv{\tilde{u}}}{\partial \tau} + S(\varphi)\frac{\nabla \varphi}{\|\nabla \varphi\|}\cdot \nabla\mv{\tilde{u}} = 0,
    \label{eq:hyperbolicExtension}
\end{equation}
which is an hyperbolic equation of Hamilton-Jacobi type. The information is propagated away from the interface along the characteristics, which are $\mv{n}$ and $-\mv{n}$. \eqref{eq:hyperbolicExtension} is discretized in space using a first-order upwind scheme and integrated in time using an Euler forward method~\cite{Peng1999}.
During the advection of the interface, the signed distance property $\Vert \nabla \varphi \Vert = 1$ is not guaranteed to be preserved. To remedy this, the level-set has to be reinitialized regularly. A high-order constrained reinitialization is used, which ensures the invariance of the zero set $\varphi=0$, i.e., the position of the interface~\cite{Hartmann2010}. The geometric properties at the interface such as the normal vector and the curvature can be calculated as
\begin{equation}
    \mv{n} = \frac{\nabla \varphi}{\Vert \nabla \varphi \Vert} 
    \quad\quad
    \kappa = \nabla \cdot \mv{n}.
\end{equation}
\subsection{Coupling conditions}
Next, the coupling of the two solution algorithms by means of boundary conditions at the two-phase interface is discussed. In the LBM algorithm, the propagation step cannot be performed at the phase boundary since no valid neighbor cell exists on the other side of the interface. This situation is illustrated in \figref{fig:coupling:lb}. To address this issue, a bounce back approach analogous to the boundary treatment at solid walls in \eqref{eq:bb} is adopted \cite{Thommes2009}. At the phase boundary the momentum source term in \eqref{eq:bb} becomes
\begin{equation}
    \Delta f_i = \frac{2 w_i}{c_s^2} (\mv{c}_i \cdot \mv{u}_I) - \frac{2 w_i}{c_s^2} \mm{\Lambda}_i q_i(1-q_i) [\mm{S}].
    \label{eq:bb_cc}
\end{equation}
The first term on the right-hand side of \eqref{eq:bb_cc} ensures the continuity at the phase interface, i.e., $[\mv{u}] = 0$, where $[\cdot]$ denotes the jump of a given quantity across the interface. The interface velocity $\mv{u}_I$ is obtained by linear interpolation between both phases. The second term accounts for the stress tensor jump at the interface, with $\mm{\Lambda}_i$ being the second-order velocity tensor. The stress jump can be decomposed into normal and tangential components
\begin{align}
  [\mm{S}] \colon \vec{n} \otimes \vec{n} &= \frac{1}{2\bar{\eta}}([p] + 2\sigma\kappa) - \frac{[\eta]}{\bar{\eta}} \colon \vec{n} \otimes \vec{n} \\
  [\mm{S}] \colon \vec{n} \otimes \vec{t}_j &= - \frac{[\eta]}{\bar{\eta}} \colon \vec{n} \otimes \vec{t}_j.
  \label{eq:Sdecomp}
\end{align}
The quantity $\Bar{\eta}$ denotes the average of the dynamic viscosity $\eta = \rho \nu$ over both phases and $\sigma$ the surface tension of the chosen material combination. The normalized wall distance $q_i$ is determined by applying the radiation theorem
\begin{equation}
    \frac{q_i}{\varphi(\mv{x})} = \frac{1-q_i}{-\varphi(\mv{x}+\mv{e}_i)}
\end{equation}
which after rearranging yields
\begin{equation}
    q_i(\mv{x})=\frac{\varphi(\mv{x})}{\varphi(\mv{x})-\varphi(\mv{x}+\mv{e}_i)}.
\end{equation}
\subsection{Initialization of newly activated fluid cells}
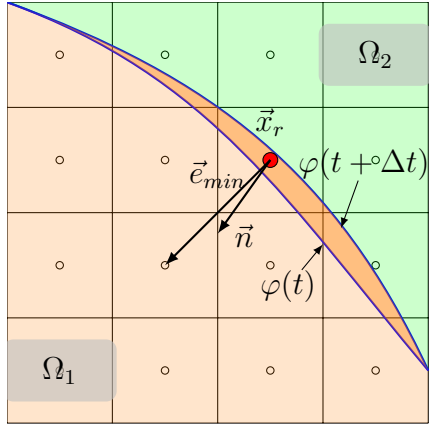
\begin{figure}
    \centering
    \begin{tikzpicture}[scale=1.4,every node/.style={minimum size=1cm}, on grid, block/.style ={rectangle, thick, fill=black!25, opacity=0.6, text width=3em, align=center, rounded corners, minimum height=2em}]
   \begin{scope}
   	    \clip (0, 0) rectangle (4,4);
       \draw[step=1, black] (0,0) grid (4,4); 
       
	   \foreach \x in {0.5,1.5,2.5,3.5}
	   \foreach \y in {0.5,1.5,2.5,3.5}
       \draw (\x,\y) circle (1pt);       
       
       \draw [thick, blue, name path=bnd] (0,4) to[out=-20,in=130] (4,0.5);
       \draw [thick, blue, name path=bnd2] (0,4) to[out=-20,in=115] (4,0.5);
       \path [thick, blue, name path=lower] (0,0) to (4,0);
       \path [thick, blue, name path=upper] (0,4) to (4,4);
       \tikzfillbetween[of=bnd and lower, on layer=]{orange, opacity=0.2};
       \tikzfillbetween[of=bnd2 and upper, on layer=]{green, opacity=0.2};
       \tikzfillbetween[of=bnd and bnd2, on layer=]{orange, opacity=0.5};
       
       \draw (0.5,0.5) node[block] (F1) {};
	   \node at (F1) {$\Omega_1$}; 
       \draw (3.5,3.5) node[block] (F2) {};
       \node at (F2) {$\Omega_2$};
       
       
       
       \draw[fill=red] (2.5, 2.5) circle (2pt) node[above, inner sep=0pt]{$\mv{x}_r$};

       \node[inner sep=0pt, minimum size=0cm] (T1) at (2.67, 1.3) {$\varphi(t)$};
       \node[inner sep=0pt, minimum size=0cm] (T2) at (3.39, 2.46) {$\varphi(t + \Delta t)$};
       \coordinate (TT1) at (3,1.7);
       \coordinate (TT2) at (3.17,1.85);
       \draw[->] (T1) -- (TT1);
       \draw[->] (T2) -- (TT2);
       \draw[->, thick] (2.5, 2.5) -- (2, 1.8) node[midway, below]{$\mv{n}$};
       \draw[->, thick] (2.5, 2.5) -- (1.5, 1.5) node[midway, above]{$\mv{e}_{min}$};
   \end{scope}
%
%
%
\end{tikzpicture}
    \caption{Example configuration illustrating the level set sweeping over a grid node $\mv{x}_r$
    ({\protect\tikz\protect\draw[fill=red] (0, 0) circle (2pt);})
    during one time step $\Delta t$ such that it becomes an active node in $\Omega_1$ and inactive in $\Omega_2$, i.e., the associated level-set value at $\varphi(\mv{x}_r)$ switches sign. The interface normal $\mv{n}$ as well as the closest lattice direction $\mv{e}_{min}$ are shown.
    }
    \label{fig:refill}
\end{figure}
During the simulation, the interface represented by the level set moves across the Eulerian grid, continually activating and deactivating fluid cells for the corresponding solver. 
While deactivated cells  require no special treatment, newly activated cells have no prior flow information and therefore must be initialized. A schematic of the problem configuration is shown in~\figref{fig:refill}.
Note that the PPDF $f_i$ representing the flow state in LB can be decomposed in its equilibrium and non-equilibrium part
\begin{equation}
    f_i = f^{eq} + f^{neq},
\end{equation}
where $f^{eq}$ encodes the macroscopic variables $\{\rho, \mv{u}\}$ according to Eqs.~(~\ref{eq:macro},~\ref{eq:feq}), while $f^{neq}$ approximates their gradients. To yield a consistent flow field after refilling newly emerged cells, the following method for both terms is proposed. First, the lattice direction closest to the interface normal $\mv{n}$ is determined by
\begin{equation}
    \mv{e}_{min}=\operatorname*{argmin}_{\mv{e}_i} \frac{\mv{e}_i\cdot\mv{n}}{\|\mv{e}_i\|}.  
\end{equation}
For the non-equilibrium part, a next neighbor approach is employed, i.e., $f^{neq}_i(\mv{x}_r)=f^{neq}_i(\mv{x_r}+\mv{e}_{min})$. The macroscopic variables are determined by interpolation between neighboring fluid nodes and information at the boundary itself
\begin{align}
    \mv{u}(\mv{x}_r)&=\frac{2}{q_i^2+3q_i+2}\mv{u}_\Gamma + \frac{2q_i}{q_i+1}\mv{u}(\mv{x}_r+\mv{e}_{min})+\frac{2q_i}{q_i+2}\mv{u}(\mv{x}_r+2\mv{e}_{min})\\
    \rho_r &=\frac{2}{q_i^2+3q_i+2}\rho_\Gamma + \frac{2q_i}{q_i+1}\rho(\mv{x}_r+\mv{e}_{min})+\frac{2q_i}{q_i+2}\rho(\mv{x}_r+2\mv{e}_{min}).
\end{align}
The boundary values $(\cdot)_\Gamma$ are based on the fluid information of the other phase and the normal component of the coupling condition described in~\eqref{eq:Sdecomp}. This enforces consistency between both phases, i.e., continuity and normal stress jump.
\subsection{Joint hierarchical Cartesian grid}
\label{sec:grid}
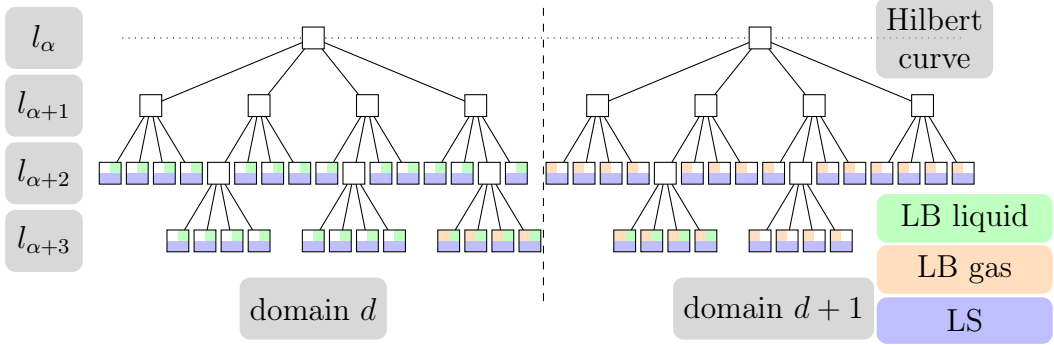
\begin{figure}[ht!]
    \centering
    \begin{tikzpicture}[scale=0.8,block/.style ={rectangle, thick, fill=black!25, opacity=0.6, text opacity=1.0, minimum width=2.5em, align=center, rounded corners, minimum height=2em}]

\tikzset{
    pics/cell/.style n args = {4}{
        code = {
        \begin{scope}[scale=0.29]
        \fill[white] (-0.5,-0.5) rectangle (0.5,0.5);
		\fill[green!25, opacity=#1] (0.0,0.0) rectangle (0.5,0.5);
		\fill[orange!25, opacity=#2] (-0.5,0.0) rectangle (0.0,0.5);
		\fill[blue!25, opacity=#3] (-0.5,-0.5) rectangle (0.5,0.0);
		\draw (-0.5,-0.5) rectangle (0.5,0.5);
		\end{scope}
        }
    }
}

\def\leveltwo{-2.5}
\def\levelthree{2*\leveltwo}
\def\levelfour{3*\leveltwo}
\def\levelfive{4*\leveltwo}

\def\fac{0.45}
\def\lvloff{-8.3}

\draw [dashed] (0.0,0.5) -- (0.0,\levelfive*\fac);
\draw [dotted] (-7,0) -- (7,0);
\node[block] at (6.5,0) {Hilbert\\curve};
\node[block] at (\lvloff,0) {$l_\alpha$};
\node[block] at (\lvloff,\leveltwo*\fac) {$l_{\alpha+1}$};
\node[block] at (\lvloff,\levelthree*\fac) {$l_{\alpha+2}$};
\node[block] at (\lvloff,\levelfour*\fac) {$l_{\alpha+3}$};

\node[block] at (-8.5*\fac,\levelfive*\fac) {domain $d$};
\node[block] at (8.5*\fac,\levelfive*\fac) {domain $d+1$};

\node[rectangle, thick, fill=green!25, opacity=1.0, text opacity=1.0, text width=5.0em, minimum width=4.0em, align=center, rounded corners, minimum height=1.5em] at (7,-3) {LB liquid};

\node[rectangle, thick, fill=orange!25, opacity=1.0, text opacity=1.0, text width=5.0em, minimum width=4.0em, align=center, rounded corners, minimum height=1.5em] at (7,-3.85) {LB gas};

\node[rectangle, thick, fill=blue!25, opacity=1.0, text opacity=1.0, text width=5.0em, minimum width=4.0em, align=center, rounded corners, minimum height=1.5em] at (7,-4.7) {LS};


\begin{scope}[scale=\fac, xshift=-8.5cm]
\path pic[local bounding box=L1] at (0,0) {cell={0}{0}{0}{0}};

\path pic[local bounding box=L11] at (-6,\leveltwo) {cell={0}{0}{0}{0}};
\path pic[local bounding box=L12] at (-2,\leveltwo) {cell={0}{0}{0}{0}};
\path pic[local bounding box=L13] at (2,\leveltwo) {cell={0}{0}{0}{0}};
\path pic[local bounding box=L14] at (6,\leveltwo) {cell={0}{0}{0}{0}};

\draw (L1) -- (L11);
\draw (L1) -- (L12);
\draw (L1) -- (L13);
\draw (L1) -- (L14);

\path pic[local bounding box=L111] at (-7.5,\levelthree) {cell={1}{0}{1}{1}};
\path pic[local bounding box=L112] at (-6.5,\levelthree) {cell={1}{0}{1}{1}};
\path pic[local bounding box=L113] at (-5.5,\levelthree) {cell={1}{0}{1}{1}};
\path pic[local bounding box=L114] at (-4.5,\levelthree) {cell={1}{0}{1}{1}};

\draw (L11) -- (L111);
\draw (L11) -- (L112);
\draw (L11) -- (L113);
\draw (L11) -- (L114);

\path pic[local bounding box=L121] at (-3.5,\levelthree) {cell={0}{0}{0}{1}};
\path pic[local bounding box=L122] at (-2.5,\levelthree) {cell={1}{0}{1}{1}};
\path pic[local bounding box=L123] at (-1.5,\levelthree) {cell={1}{0}{1}{1}};
\path pic[local bounding box=L124] at (-0.5,\levelthree) {cell={1}{0}{1}{1}};

\draw (L12) -- (L121);
\draw (L12) -- (L122);
\draw (L12) -- (L123);
\draw (L12) -- (L124);

\path pic[local bounding box=L131] at (0.5,\levelthree) {cell={1}{0}{1}{1}};
\path pic[local bounding box=L132] at (1.5,\levelthree) {cell={0}{0}{0}{1}};
\path pic[local bounding box=L133] at (2.5,\levelthree) {cell={1}{0}{1}{1}};
\path pic[local bounding box=L134] at (3.5,\levelthree) {cell={1}{0}{1}{1}};

\draw (L13) -- (L131);
\draw (L13) -- (L132);
\draw (L13) -- (L133);
\draw (L13) -- (L134);

\path pic[local bounding box=L141] at (4.5,\levelthree) {cell={1}{0}{1}{1}};
\path pic[local bounding box=L142] at (5.5,\levelthree) {cell={1}{0}{1}{1}};
\path pic[local bounding box=L143] at (6.5,\levelthree) {cell={0}{0}{0}{1}};
\path pic[local bounding box=L144] at (7.5,\levelthree) {cell={1}{0}{1}{1}};

\draw (L14) -- (L141);
\draw (L14) -- (L142);
\draw (L14) -- (L143);
\draw (L14) -- (L144);

\path pic[local bounding box=L1211] at (-2,\levelfour) {cell={1}{0}{1}{0}};
\path pic[local bounding box=L1212] at (-3,\levelfour) {cell={1}{0}{1}{0}};
\path pic[local bounding box=L1213] at (-4,\levelfour) {cell={1}{0}{1}{0}};
\path pic[local bounding box=L1214] at (-5,\levelfour) {cell={1}{0}{1}{0}};

\draw (L121) -- (L1211);
\draw (L121) -- (L1212);
\draw (L121) -- (L1213);
\draw (L121) -- (L1214);

\path pic[local bounding box=L1321] at (0,\levelfour) {cell={1}{0}{1}{0}};
\path pic[local bounding box=L1322] at (1,\levelfour) {cell={1}{0}{1}{0}};
\path pic[local bounding box=L1323] at (2,\levelfour) {cell={1}{0}{1}{0}};
\path pic[local bounding box=L1324] at (3,\levelfour) {cell={1}{0}{1}{0}};

\draw (L132) -- (L1321);
\draw (L132) -- (L1322);
\draw (L132) -- (L1323);
\draw (L132) -- (L1324);

\path pic[local bounding box=L1431] at (5,\levelfour) {cell={1}{1}{1}{0}};
\path pic[local bounding box=L1432] at (6,\levelfour) {cell={1}{1}{1}{0}};
\path pic[local bounding box=L1433] at (7,\levelfour) {cell={1}{1}{1}{0}};
\path pic[local bounding box=L1434] at (8,\levelfour) {cell={1}{1}{1}{0}};

\draw (L143) -- (L1431);
\draw (L143) -- (L1432);
\draw (L143) -- (L1433);
\draw (L143) -- (L1434);

\end{scope}

\begin{scope}[scale=\fac, xshift=8.0cm]
\path pic[local bounding box=L1] at (0,0) {cell={0}{0}{0}{0}};

\path pic[local bounding box=L11] at (-6,\leveltwo) {cell={0}{0}{0}{0}};
\path pic[local bounding box=L12] at (-2,\leveltwo) {cell={0}{0}{0}{0}};
\path pic[local bounding box=L13] at (2,\leveltwo) {cell={0}{0}{0}{0}};
\path pic[local bounding box=L14] at (6,\leveltwo) {cell={0}{0}{0}{0}};

\draw (L1) -- (L11);
\draw (L1) -- (L12);
\draw (L1) -- (L13);
\draw (L1) -- (L14);

\path pic[local bounding box=L111] at (-7.5,\levelthree) {cell={0}{1}{1}{1}};
\path pic[local bounding box=L112] at (-6.5,\levelthree) {cell={0}{1}{1}{1}};
\path pic[local bounding box=L113] at (-5.5,\levelthree) {cell={0}{1}{1}{1}};
\path pic[local bounding box=L114] at (-4.5,\levelthree) {cell={0}{1}{1}{1}};

\draw (L11) -- (L111);
\draw (L11) -- (L112);
\draw (L11) -- (L113);
\draw (L11) -- (L114);

\path pic[local bounding box=L121] at (-3.5,\levelthree) {cell={0}{0}{0}{1}};
\path pic[local bounding box=L122] at (-2.5,\levelthree) {cell={0}{1}{1}{1}};
\path pic[local bounding box=L123] at (-1.5,\levelthree) {cell={0}{1}{1}{1}};
\path pic[local bounding box=L124] at (-0.5,\levelthree) {cell={0}{1}{1}{1}};

\draw (L12) -- (L121);
\draw (L12) -- (L122);
\draw (L12) -- (L123);
\draw (L12) -- (L124);

\path pic[local bounding box=L131] at (0.5,\levelthree) {cell={0}{1}{1}{1}};
\path pic[local bounding box=L132] at (1.5,\levelthree) {cell={0}{0}{0}{1}};
\path pic[local bounding box=L133] at (2.5,\levelthree) {cell={0}{1}{1}{1}};
\path pic[local bounding box=L134] at (3.5,\levelthree) {cell={0}{1}{1}{1}};

\draw (L13) -- (L131);
\draw (L13) -- (L132);
\draw (L13) -- (L133);
\draw (L13) -- (L134);

\path pic[local bounding box=L141] at (4.5,\levelthree) {cell={0}{1}{1}{1}};
\path pic[local bounding box=L142] at (5.5,\levelthree) {cell={0}{1}{1}{1}};
\path pic[local bounding box=L143] at (6.5,\levelthree) {cell={0}{1}{1}{1}};
\path pic[local bounding box=L144] at (7.5,\levelthree) {cell={0}{1}{1}{1}};

\draw (L14) -- (L141);
\draw (L14) -- (L142);
\draw (L14) -- (L143);
\draw (L14) -- (L144);

\path pic[local bounding box=L1211] at (-2,\levelfour) {cell={1}{1}{1}{0}};
\path pic[local bounding box=L1212] at (-3,\levelfour) {cell={1}{1}{1}{0}};
\path pic[local bounding box=L1213] at (-4,\levelfour) {cell={1}{1}{1}{0}};
\path pic[local bounding box=L1214] at (-5,\levelfour) {cell={1}{1}{1}{0}};

\draw (L121) -- (L1211);
\draw (L121) -- (L1212);
\draw (L121) -- (L1213);
\draw (L121) -- (L1214);

\path pic[local bounding box=L1321] at (0,\levelfour) {cell={0}{1}{1}{0}};
\path pic[local bounding box=L1322] at (1,\levelfour) {cell={0}{1}{1}{0}};
\path pic[local bounding box=L1323] at (2,\levelfour) {cell={0}{1}{1}{0}};
\path pic[local bounding box=L1324] at (3,\levelfour) {cell={0}{1}{1}{0}};

\draw (L132) -- (L1321);
\draw (L132) -- (L1322);
\draw (L132) -- (L1323);
\draw (L132) -- (L1324);

\end{scope}
\end{tikzpicture}
    \caption{Example cutout of a parallel quad tree data structure for a two-dimensional hierarchical Cartesian grid. The LS solver is active on all leaf-cells
    \protect\tikz{\protect\fill[white] (-0.15,-0.15) rectangle (0.15,0.15);
	   \protect\fill[blue!25] (-0.15,-0.15) rectangle (0.15,0.0);
	   \protect\draw (-0.15,-0.15) rectangle (0.15,0.15);}, i.e., cells on the maximum refinement level, while both flow solvers are only active together in the cells where the interface is present
    \protect\tikz{\protect\fill[white] (-0.15,-0.15) rectangle (0.15,0.15);
	   \protect\fill[blue!25] (-0.15,-0.15) rectangle (0.15,0.0);
       \protect\fill[green!25] (0.0,0.0) rectangle (0.15,0.15);
		\protect\fill[orange!25] (-0.15,0.0) rectangle (0.0,0.15);
	   \protect\draw (-0.15,-0.15) rectangle (0.15,0.15);}.
    In the bulk of the respective fluid domains, only the corrsesponding LB gas or LB liquid solver is active
    \protect\tikz{\protect\fill[white] (-0.15,-0.15) rectangle (0.15,0.15);
		\protect\fill[orange!25] (-0.15,0.0) rectangle (0.0,0.15);
	   \protect\draw (-0.15,-0.15) rectangle (0.15,0.15);}/\protect\tikz{\protect\fill[white] (-0.15,-0.15) rectangle (0.15,0.15);
       \protect\fill[green!25] (0.0,0.0) rectangle (0.15,0.15);
	   \protect\draw (-0.15,-0.15) rectangle (0.15,0.15);}
. Here, two subdomains for the parallelization as shown.}
    \label{fig:tree}
\end{figure}

One of the key features of m-AIA is the fully automatic grid generation on HPC hardware~\cite{LintermannGrid}. For the grid generation as well as the simulation itself, the grid is represented using a quad-/octree data structure to encode the hierarchical relationship between coarse and fine cells. In parallel simulations, the global tree is partitioned on a user-defined level $l_\alpha$, on which a Hilbert curve is used to transform the data structure into a one-dimensional partitioning problem. After partitioning, each subdomain is assigned a forest of subtrees. Each solver, operates on a subset of the hierarchical Cartesian grid. This allows for different refinement specifications for each solver as well as zonal methods, where different solvers are active in different regions of the computational domain. For liquid-gas multiphase flows, where each fluid phase is handled by a different solution algorithm and a third solver captures the phase interface, an example grid data structure is shown in~\figref{fig:tree}. One of the major advantages of the joint grid is the efficient exchange of coupling terms for surface or volume coupled problems without the additional need for communication between solvers.

\subsection{Adaptive mesh refinements}

Adaptive mesh refinement (AMR) is used to improve the accuracy of simulations by dynamically refining the computational mesh in regions where complex physical phenomena or steep gradients occur, while keeping the mesh coarser in less critical areas. This selective refinement allows for efficient use of computational resources, as it achieves higher resolution only where it is most needed, rather than uniformly across the entire domain~\cite{Berger1984}. The solver framework m-AIA provides AMR via the joint hierarchical Cartesian grid~\cite{Schlimpert2016, Hartmann2008}. The refinement and coarsening of cells is controlled by solution dependent sensor functions, which are defined by the different solvers. In general, a sensor function is defined as
\begin{equation}
    s(\mv{x})=
    \begin{cases}
        -1,\quad\text{tagged for coarsening}\\
        ~~~0,\quad\text{unchanged}\\
        ~~~1,\quad\text{tagged for refinement}
    \end{cases}
\end{equation}
Note that this approach allows the refinement of one solver based on sensor functions provided by another solver, thus offering maximum flexibility for coupled simulation.\\
In summary, the presented method introduces a fully coupled two-fluid approach that avoids non-local regularization by employing independent, phase-specific solution algorithms—such as a hybrid of different Lattice Boltzmann or Finite Volume schemes—optimized for the unique physical properties of each phase. These solvers are integrated via a joint hierarchical Cartesian grid, which eliminates inter-process communication by enabling direct memory exchange of coupling data and supports localized adaptive mesh refinement. The interface is captured using a level-set method, where physical jump conditions (surface tension and viscosity) are enforced through a modified bounce-back boundary condition, and newly activated cells are initialized using a novel non-equilibrium refilling strategy that ensures inter-phase consistency.
\section{Results}
\label{sec:results}
The results section is structured as follows. First, we validate the approach by simulating shear-dominated processes using laminar stratified flows. Next, the complexity is increased by introducing gravity and surface tension. Two cases of single rising bubbles under gravity in three dimensions are simulated. For one case, the effectiveness of adaptive mesh refinement is evaluated. Finally, the method's capability to handle multiple interacting bubbles is demonstrated by simulating a cluster of nine rising bubbles under gravity.
\subsection{Laminar stratified flows}

To  validate the proper coupling of the tangential stresses between both fluid solvers, two cases of laminar stratified flow, i.e., laminar Couette and Poiseuille flow, are considered, both of which have analytic solutions. 
The two-dimensional computational domain is $[0, 2]\times[-1,1]$, where $y<0$ is filled by fluid $1$ and $y\geq0$ by fluid $2$. While the lower wall at $y=-1$ is stationary, the upper wall at $y=1$ is moving with $u_W$. In the streamwise $x$-direction, a periodic boundary condition is applied. The analytical solution for the Couette flow is given by the piecewise linear equation
\begin{equation}
    u(y)/u_W= 
\begin{cases}
    & \frac{\eta_1}{\eta_1+\eta_2}y+\frac{\eta_2}{\eta_1+\eta_2},\quad y\geq0,\\
    & \frac{\eta_2}{\eta_1+\eta_2}y+\frac{\eta_2}{\eta_1+\eta_2},\quad y<0.
\end{cases}
\end{equation}
For the Poiseuille flow, the same domain is used, with both walls being stationary. The flow is driven by a constant volume forcing $\del{p}{x}$. The analytical solution for this case is given by
\begin{equation}
    u(y)/\hat{u}= 
\begin{cases}
    -\frac{\eta_1}{\eta_2}(y^2-\frac{\eta_1-\eta_2}{\eta_1+\eta_2}y-\frac{2\eta_2}{\eta_1+\eta_2}),&\quad y\geq0,\\
    -(y^2-\frac{\eta_1-\eta_2}{\eta_1+\eta_2}y-\frac{2\eta_1}{\eta_1+\eta_2}),&\quad y<0,
\end{cases}
\end{equation}
with $\hat{u}=-\frac{1}{2\eta_1}\del{p}{x}$.
\begin{figure}[ht!]
    \centering
    \begin{subfigure}[t]{0.49\textwidth}
    \centering
    \includegraphics[width=1\textwidth]{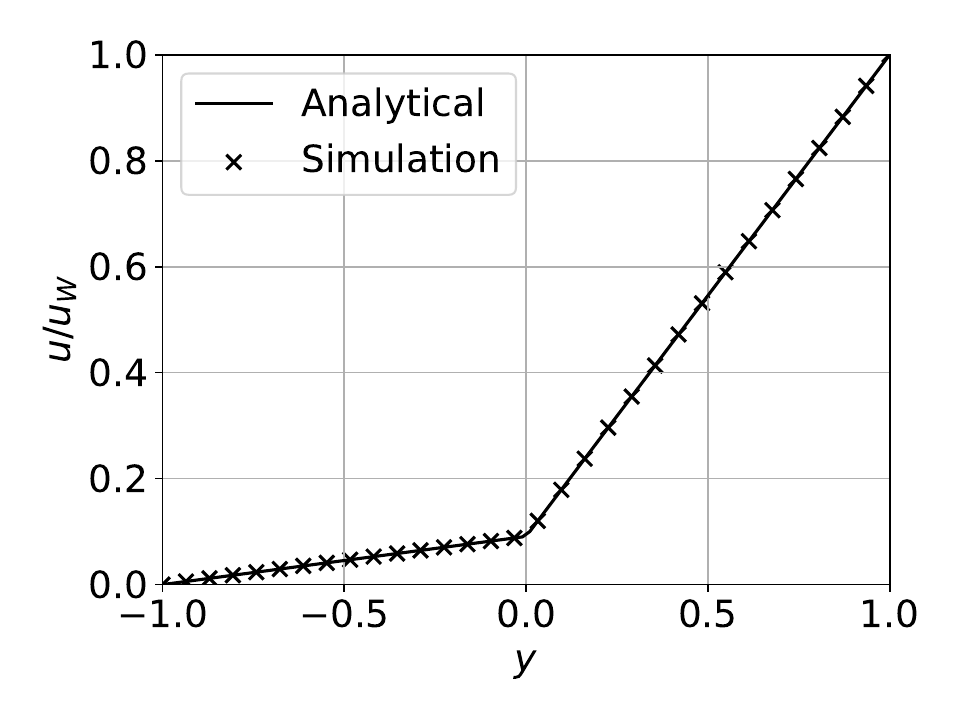}
    \caption{Couette flow}
    \end{subfigure}
    \begin{subfigure}[t]{0.49\textwidth}
    \centering
    \includegraphics[width=1\textwidth]{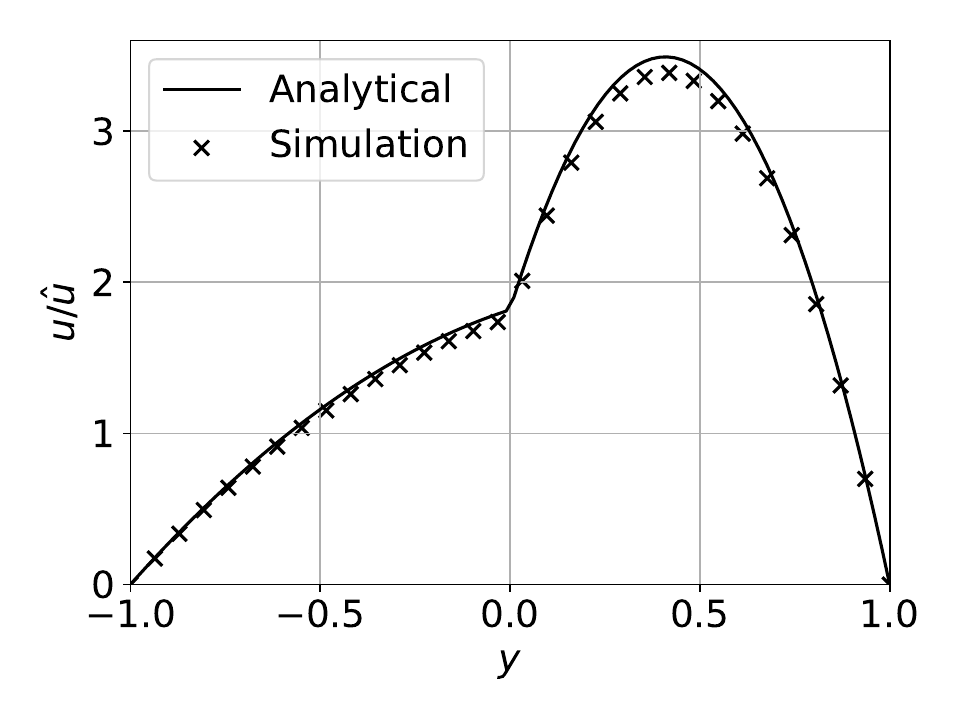}
    \caption{Poiseuille flow}
    \end{subfigure}
    \caption{Velocity profiles for different stratified two-phase flows for a viscosity ratio $\eta_1/\eta_2=10$.}
    \label{fig:strati}
\end{figure}
The simulation results for the viscosity ratio $\eta_1/\eta_2=10$ compared to the analytical solutions are shown in~\figref{fig:strati}. In both cases, the velocity profile is captured accurately. The Poiseuille flow has a maximum deviation of $\|\frac{u-u_{ref}}{u_{ref}}\|_\infty=0.03$.

\subsection{Single rising bubble}

\begin{figure}[ht!]
    \centering
    \begin{tikzpicture}[scale=0.85,every node/.style={minimum size=0.5cm}, on grid, block/.style ={rectangle, thick, fill=black!25, opacity=0.6, text width=2.2em, align=center, rounded corners, minimum height=1.2em}]

\begin{scope}
    
    \shade[ball color = gray!40, opacity = 0.4] (2,2,-2) circle (1cm);
    \draw (2,2,-2) circle (1cm);
    \fill[fill=black] (0,0) circle (1pt);
    

    \pgfmathsetmacro{\cubex}{4}
    \pgfmathsetmacro{\cubey}{8}
    \pgfmathsetmacro{\cubez}{4}
    \draw [draw=black, every edge/.append style={draw=blue, densely dashed, opacity=.5}]
    (4,0,0) coordinate (o) -- ++(-\cubex,0,0) coordinate (a) -- ++(0,\cubey,0) coordinate (b) edge coordinate [pos=1] (g) ++(0,0,-\cubez)  -- ++(\cubex,0,0) coordinate (c) -- cycle
    (o) -- ++(0,0,-\cubez) coordinate (d) -- ++(0,\cubey,0) coordinate (e) edge (g) -- (c) -- cycle
    (o) -- (a) -- ++(0,0,-\cubez) coordinate (f) edge (g) -- (d) -- cycle;

    \draw[ell] (1,2,-2) --++ (2,0) node[midway, below] {$d_B$};
    \draw[ell] (0,-0.2) --++ (4,0) node[midway, below] {$2d_B$};
    \draw[ell] (4,-0.2) --++ (0,0,-4) node[midway, below right, inner sep=0] {$2d_B$};
    \draw[ell] (-0.2,0) --++ (0,8) node[midway, left] {$4d_B$};

    \draw (2,6,-2) node (sbc) {};
    \node at (sbc) {$\Gamma_{slip}$};
    \draw (0,6,-2) --++ (1.5,0);
    \draw (4,6,-2) --++ (-1.5,0);
    \draw (2,6,0) --++ (0,0,-1.5);
    \draw (2,6,-4) --++ (0,0,+1.5);
    \draw (2,0,-2) --++ (0, 0.2) node[below] {$\Gamma_{no\,slip}$};
    \draw (2,8,-2) --++ (0, -0.2) node[above] {$\Gamma_{no\,slip}$};
    \draw (2, 3, -2) --++ (0, 0.2, 0) node[above] {$\Gamma_{interface}$};
       
    \draw (2,2.5,-2) node[block] (F1a) {};
    \node at (F1a) {gas};
    \draw (2.0,4.5,-2) node[block] (F2) {};
    \node at (F2) {liquid};
\end{scope}
\end{tikzpicture}
    \caption{Computational setup for a single bubble rising under gravity in three dimensions.}
    \label{fig:comp_setup}
\end{figure}
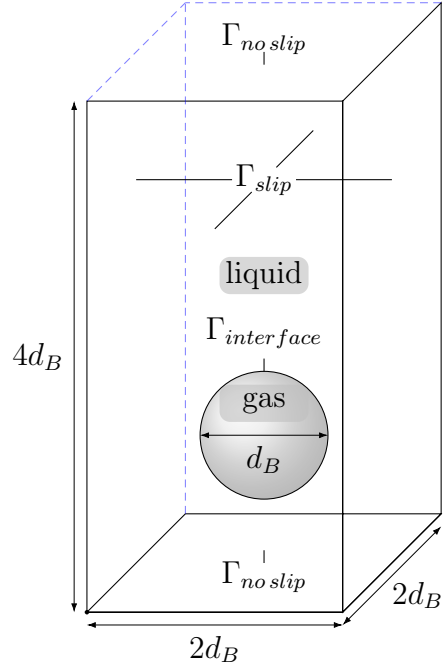

To further validate the proposed method, the canonical benchmarks of Safi et al.~\cite{Safi3D} are simulated. The reference solution was obtained using a coupled phase-field-lattice Boltzmann solver in three dimensions, following the ideas of \cite{Fakhari2016}. Therefore, the reference is using a one-fluid approach. A bubble of diameter $d_B$ is placed in a closed container of width $2d_B$ and height $4d_B$. At the top and bottom walls, a no-slip boundary condition is applied, i.e., $\mv{u}_\Gamma=0$, while a slip boundary condition is applied at the lateral walls, i.e, $\mv{u}_\Gamma\cdot\mv{n}_\Gamma=0$ and $\nabla \mv{u}_\Gamma \cdot \mv{n}_\Gamma=0$. The computational setups are shown in~\figref{fig:comp_setup}.
%
%
\begin{table}[ht!]
    \centering
    \begin{tabular}{lcccccccccc}
        \toprule
        Test case & $Re$ & $Eo$ & $\rho_1/\rho_2$ & $\eta_1/\eta_2$ \\
        \midrule
        I & 35 & 10 & 10 & 10 \\
        II & 35 & 125 & 1000 & 100 \\
        \bottomrule
    \end{tabular}
    \caption{Parameters for the single rising bubble benchmark taken from~\cite{Hysing2D, Safi3D}.}
    \label{tab:parameters}
\end{table}
Besides the density ratio and the viscosity ratio, the problem is characterized by two additional dimensionless numbers, i.e., the Reynolds ($Re$) and Eötvös ($Eo$) numbers, which are defined by
\begin{align*}
    Re &= \frac{\rho_l \sqrt{gd_b^3}}{\eta_l}\\
    Eo &= \frac{\rho_l g d_b^2}{\sigma}.
\end{align*}
Following~\cite{Hysing2D, Safi3D}, two parameter sets specified in table Tab.~\ref{tab:parameters} are considered.

To compare the results, several integral quantities are introduced. The center of gravity $\mv{x}_c$ and the velocity of the bubble $\mv{u}_b$ are
\begin{equation}
  \mv{x}_b=\frac{\int_{\Omega_{gas}}{\mv{x}dV}}{\int_{\Omega_{gas}}dV} \quad\quad
  \mv{u}_b=\frac{\int_{\Omega_{gas}}{\mv{u}dV}}{\int_{\Omega_{gas}}dV}.
\end{equation}
The integrals of type $\int_{\Omega_{gas}}\mv{f}dV$ are approximated by $\int_{\Omega}\mv{f}\heaviside(\varphi)dV$, where $\heaviside$ is the Heaviside function.
%
%
%
%
%
To compare the current results with the reference data~\cite{Hysing2D, Safi3D}, the LB time is normalized using
\begin{equation}
    t_{ref}=\sqrt{\frac{g^*}{gL^*}},
\end{equation}
where $g^*$ is the gravity in lattice units, $g$ the gravity in reference units, and $L^*$ the width of the domain, i.e., $2d_B$.\\
%
%
\begin{figure}[ht!]
    \centering
    \begin{subfigure}[t]{1.0\textwidth}
    \centering
    \includegraphics[width=0.49\textwidth]{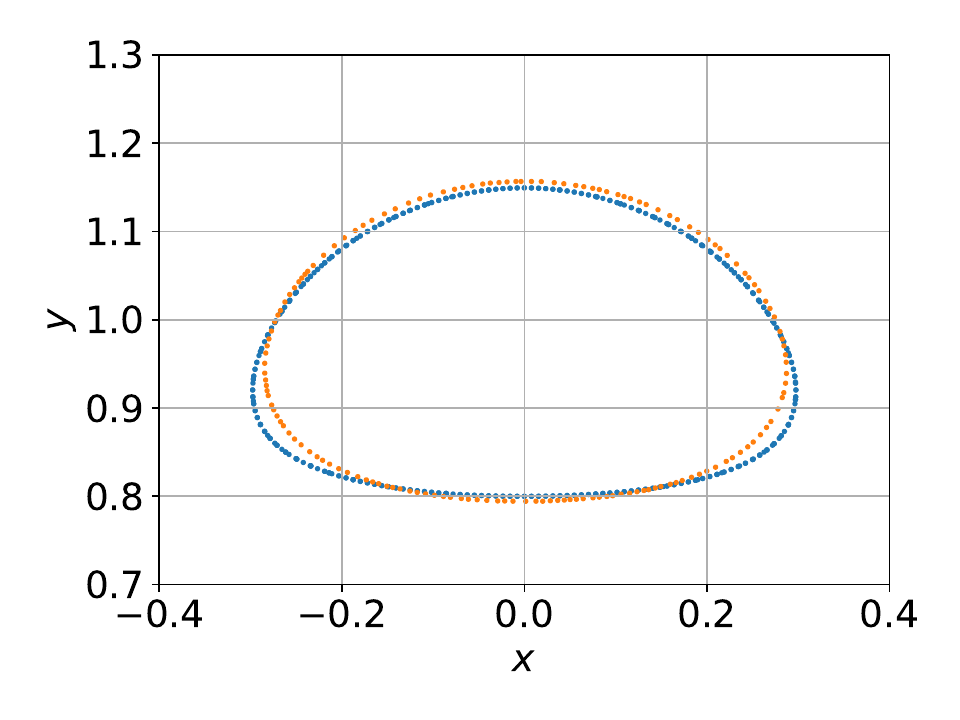}
    \includegraphics[width=0.49\textwidth]{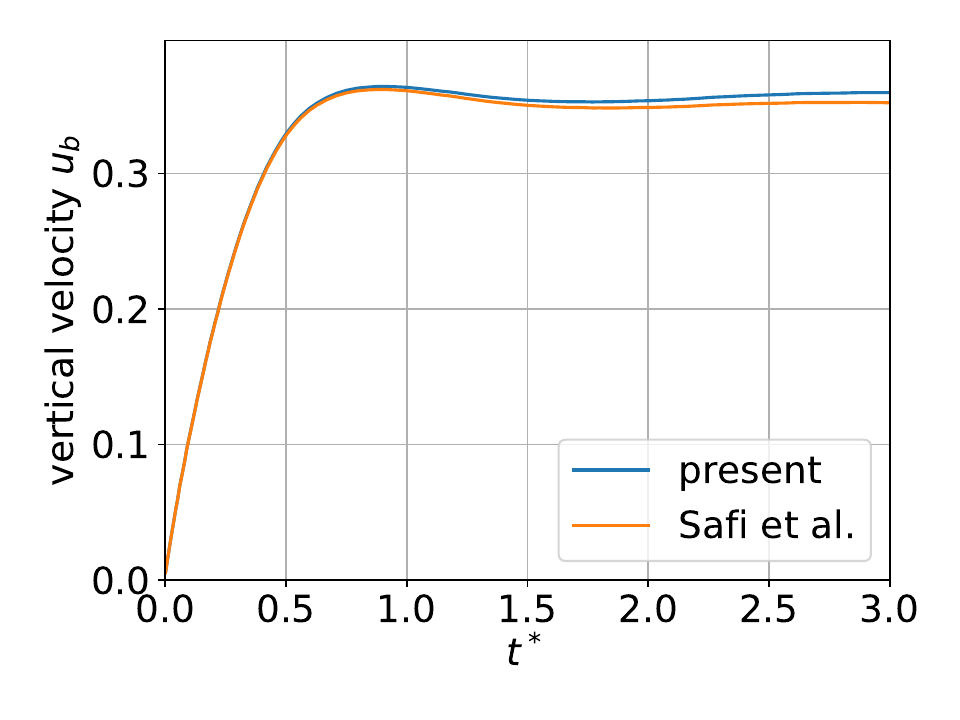}
    \subcaption{Case I}
    \label{fig:3DI}
    \end{subfigure}
    \begin{subfigure}[t]{1.0\textwidth}
    \centering
    \includegraphics[width=0.49\textwidth]{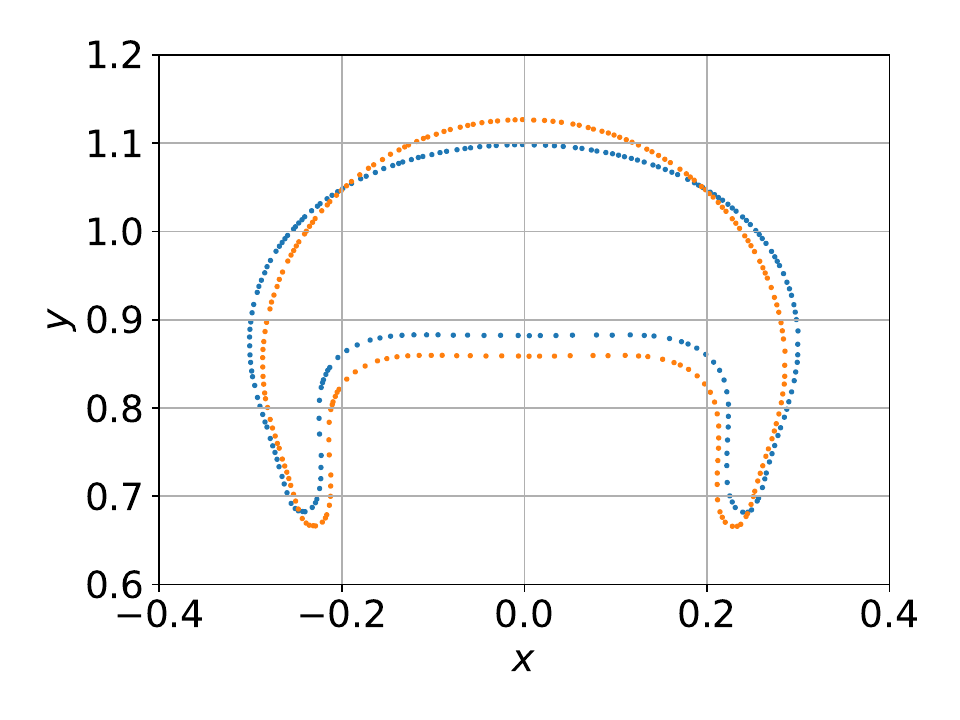}
    \includegraphics[width=0.49\textwidth]{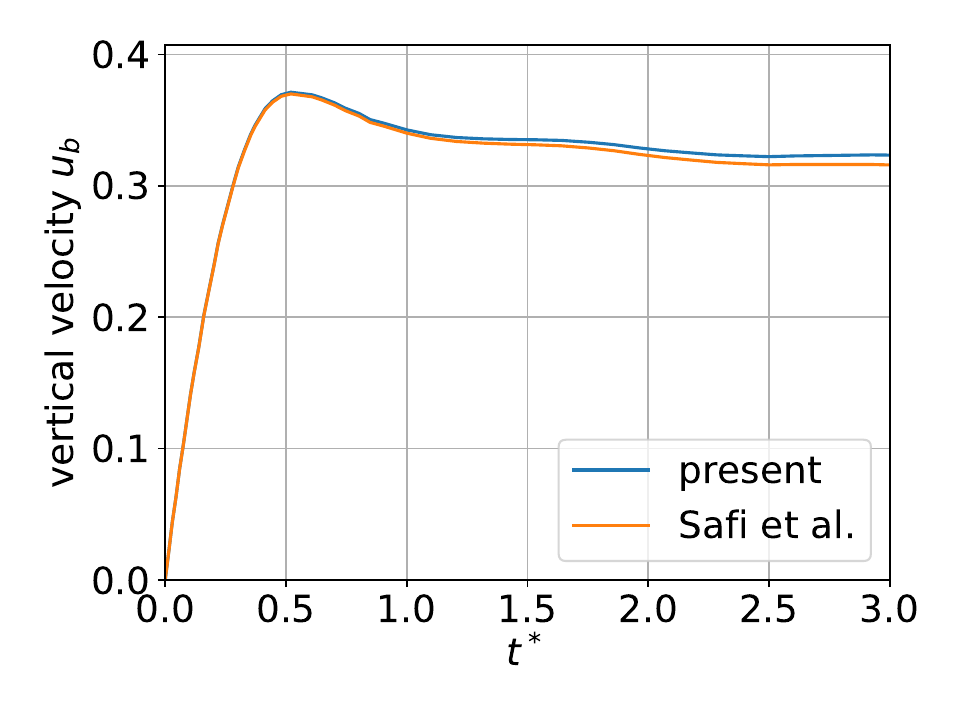}
    \subcaption{Case II}
    \label{fig:3DII}
    \end{subfigure}
    \caption{Bubble shape at $t^*=1.5$ and velocity for benchmark case I and II in three dimensions. Note, the shapes are shown with matching center of gravity.}
    \label{fig:3D}
\end{figure}
The results for the two parameter sets (cases I and II) in three dimensions are shown in Fig.~\ref{fig:3D}. In the corresponding simulations, a BGK collision kernel was used for the liquid phase, while the cumulant model was employed for the gas phase. Due to the confined nature of the setup ($L^*=2d_B$), the remaining compressibility in the liquid flow solver plays an important role. Here, the results were obtained for the Mach number $Ma=0.02$, which was the lowest stable $Ma$ possible using the BGK collision kernel.\\

\subsubsection{Uniform mesh simulation}
All results are obtained using a uniform grid with $d_B/\Delta x=64$.
\begin{figure}[ht!]
    \centering
    \includegraphics[clip, trim=15cm 0cm 10cm 0cm, width=0.7\linewidth]{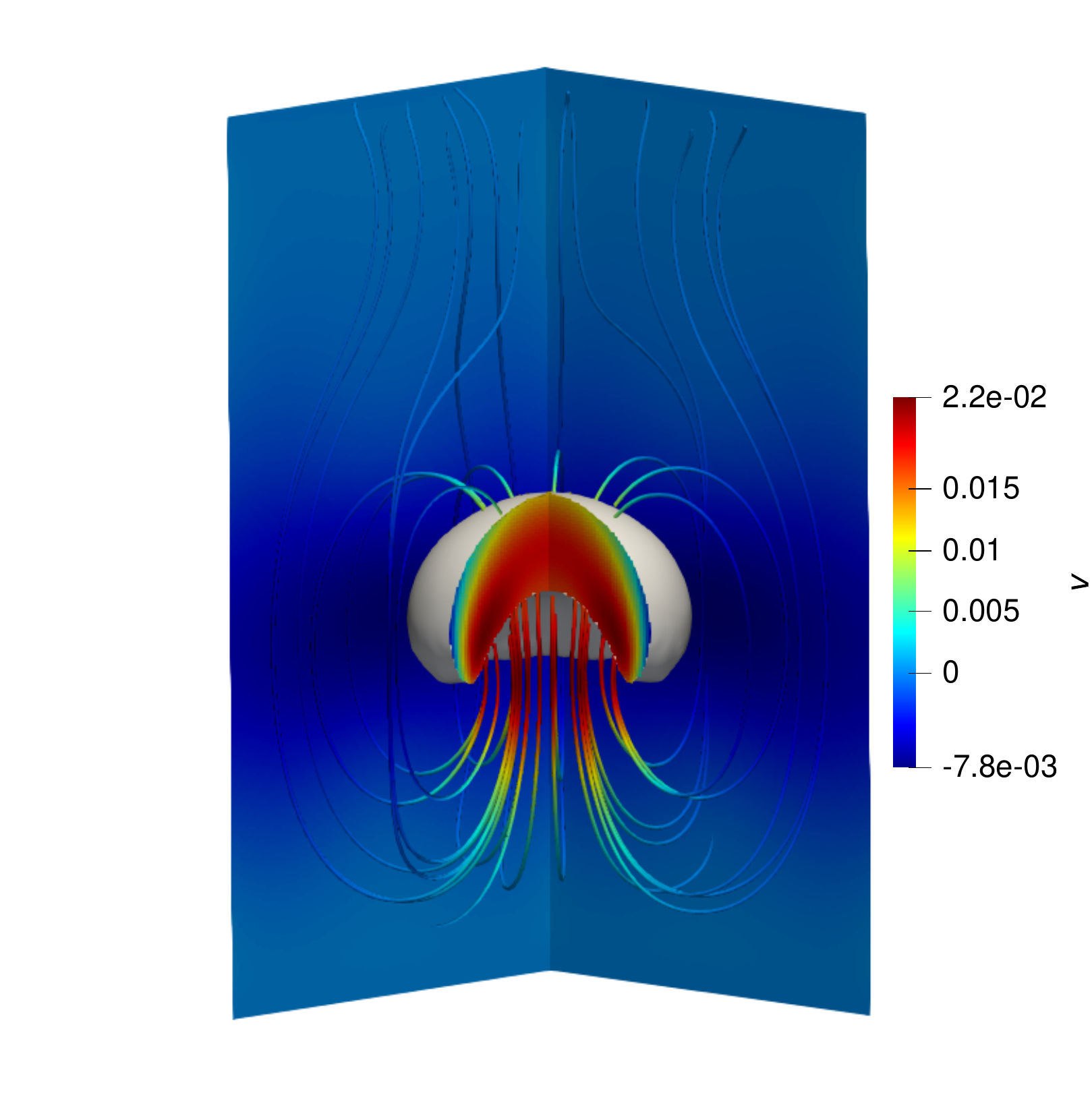}
    \caption{Snapshot of the vertical velocity component for case 3D-II at $t^*=1$. The velocity is shown for the gas and the liquid phase. The opaque gray surface defines the bubble shape.}
    \label{fig:vel_3DII}
\end{figure}
In general, the rising velocity is in good agreement with the literature, i.e., the deviation is less than $5\%$. Note that the shapes in the left columns are aligned by their centers of gravity to separate translational differences from variations in the bubble outlines. The bubble shape is captured adequately for the low $Eo$ number case, i.e., 3D-I Fig.~\ref{fig:3DI}~(left). For the high $Eo$ number case, 3D-II Fig.~\ref{fig:3DII}~(left), some deviations from the references are evident.
For the case 3D-II in~\figref{fig:3DII}~(left), the skirt shape is captured reasonably close, while the top part shows slight deviations.

A snapshot of the interior and exterior flow fields of the case 3D-II is shown in~\figref{fig:vel_3DII}. Here, the interplay between the vorticity generated by the bubble movement and the deformation following that vorticity can be seen.
\subsubsection{Adaptive mesh refinement}
\begin{figure}[ht!]
    \centering
    \begin{subfigure}[t]{0.4\textwidth}
    \vskip 8pt
    \begin{tikzpicture}
    \node[anchor=south west,inner sep=0] (image) at (0,0) {\includegraphics[width=1.08\linewidth]{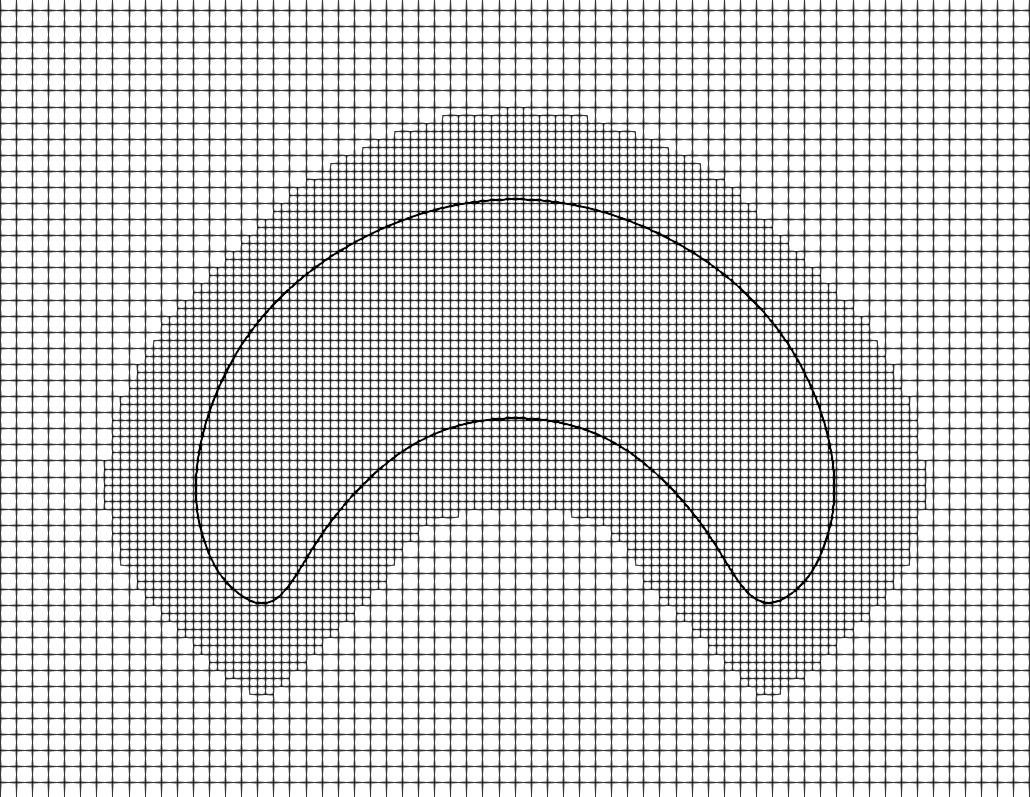}};
    \begin{scope}[x={(image.south east)},y={(image.north west)}]
    \draw[ell] (0.5,0.75) --++ (0.0,0.116) node[midway, right] {$d_{ref}$};
    \node (l) at (0.25, 0.85) {$liquid$};
    \node (l) at (0.5, 0.6) {$gas$};
    \node (A) at (0.06,0.9) {a)};
    \end{scope}
    \end{tikzpicture}%
    \vskip 19pt
    \subcaption[]{}
    \label{fig:adap_grid}
    \end{subfigure}
    \begin{subfigure}[t]{0.59\textwidth}
    \vskip 0pt
    \begin{tikzpicture}
    \node[anchor=south west,inner  sep=0] (image) at (0,0) {\includegraphics[width=0.97\linewidth]{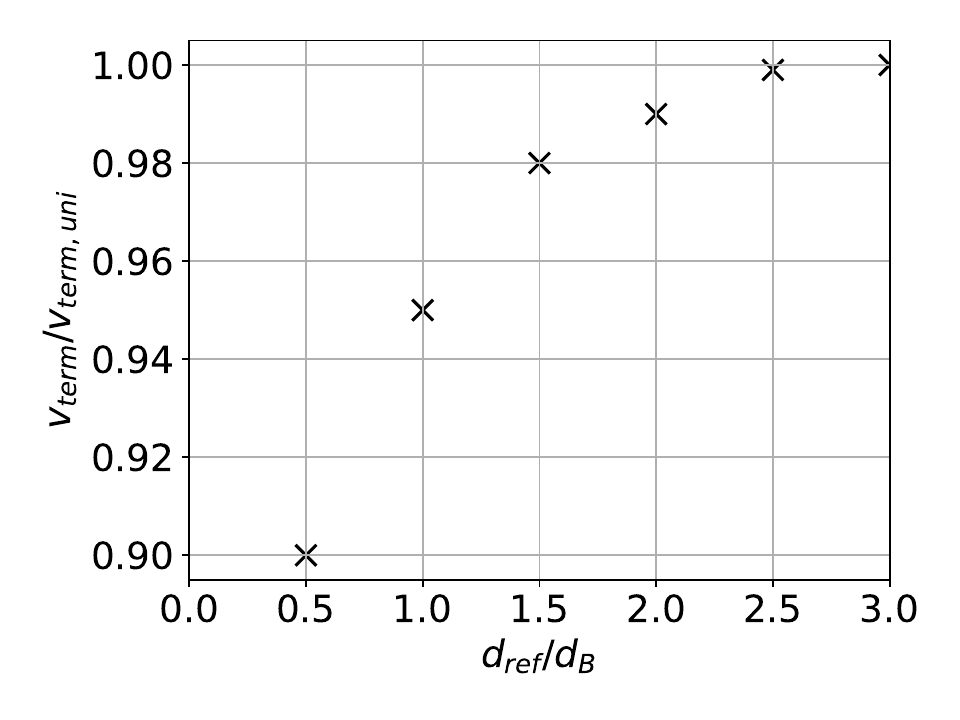}};
    \begin{scope}[x={(image.south east)},y={(image.north west)}]
    \node (A) at (0.25,0.86) {b)};
    \end{scope}
    \end{tikzpicture}%
    \vskip -10pt
    \subcaption[]{}
    \label{fig:adap_res}
    \end{subfigure}
    \caption{Two-dimensional slice of the adaptively refined mesh. The refined distance in the liquid domain is denoted with $d_{ref}$ (a). Terminal velocity for case 3D-II using adaptive mesh refinement with different refinement distances (b).}
    \label{fig:adap_setup}
\end{figure}
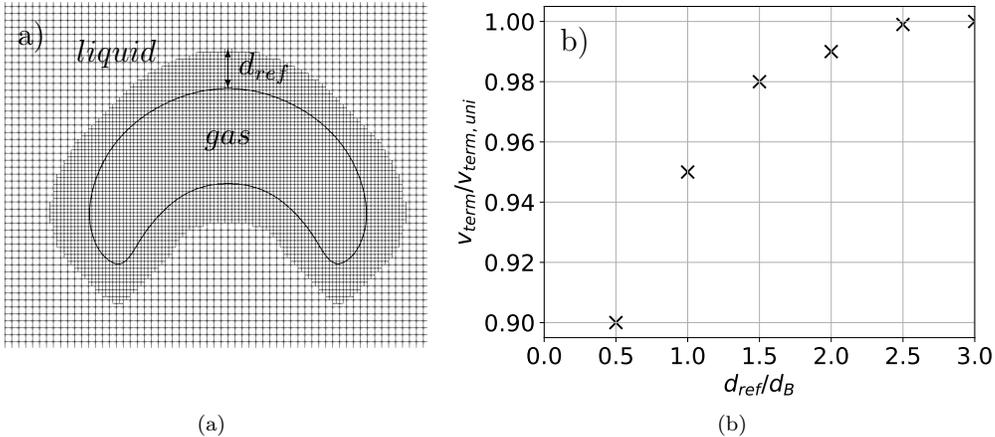

Next, the three-dimensional case II is analyzed using adaptive mesh refinement (AMR). This case is chosen since it is numerically challenging and sufficiently large to benefit from AMR. The uniform background grid is chosen as $(d_B/\Delta x)_{base}=32$ and a local adaptive refinement as $(d_B/\Delta x)_{ref}=64$, i.e., the fine grid size corresponds to the uniform grid of the previous section. A simple sensor function based on the level set is used to refine a defined region around the bubble surface. This way the deformable interface, the near-field pressure distribution, and the vorticity are resolved on the finest level, while the wake and far field are not highly resolved. The sensor function is given by
\begin{equation}
 s_{LS}(\mv{x})= 
 \begin{cases}
 -1,\quad \varphi(\mv{x}) > d_{ref} \land l(\mv{x}) = l_{ref}\\
        ~~~1,\quad \varphi(\mv{x}) \leq d_{ref} \land l(\mv{x}) = l_{base}\\
        ~~~0,\quad\text{otherwise.}
\end{cases}
\end{equation}
As shown in~\figref{fig:adap_grid}, setting the refinement scale $d_{ref}=0$ results in a grid which is refined in the gas phase and coarse in the liquid phase, and $d_{ref}\rightarrow\infty$ is analogous to the uniform fine grid. Note that the current approach is limited to phase interfaces with the same level of refinement on both sides. 
The terminal bubble velocity for different refinement distances $d_{ref}=[0.0,3.0]$ is shown in~\figref{fig:adap_grid}. For the smallest refinement distance of $d_{ref}/d_B=0.5$ the deviation to the uniform case is about $10\%$. This discrepancy is seen to continuously decrease with increasing $d_{ref}$ and drops below $1\%$ as the refinement region reaches $d_{ref}/d_B=2.5$. That is, to yield a reasonable accuracy of below $1\%$, a refined volume $\approx2.6 \pi d_B^3$ is needed.
\subsection{Cluster of rising bubbles}
\begin{figure}[ht!]
    \centering
    \includegraphics[clip, trim=8cm 0cm 12cm 0cm, width=1.0\linewidth]{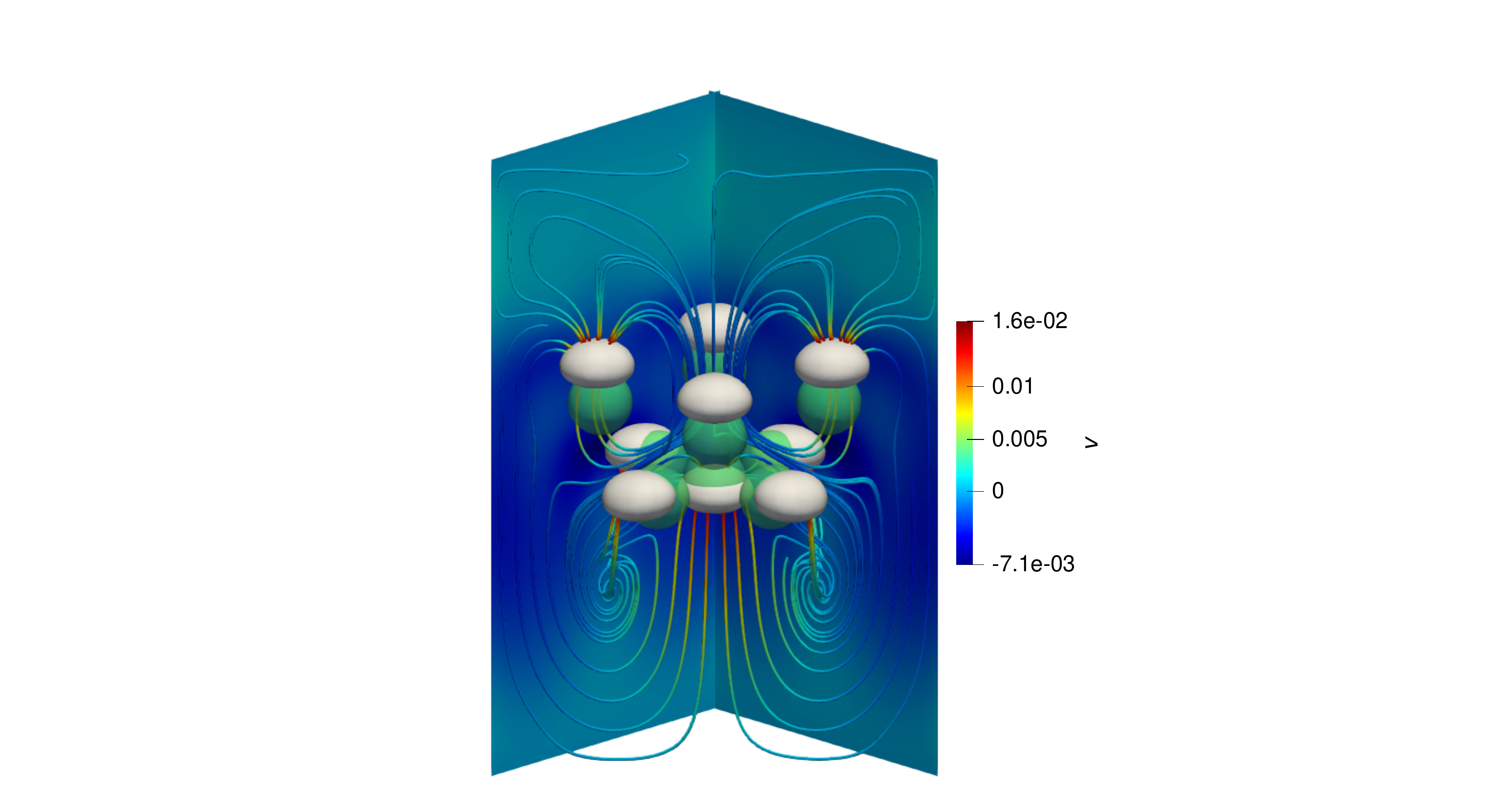}
    \caption{Snapshot of the vertical velocity component of 9 rising bubbles at $t^*=1$. The bubble shapes are shown as opaque surfaces. The initial bubble configuration is shown as lucid surfaces.}
    \label{fig:vel_3D9}
\end{figure}
A cluster of bubbles rising under gravity is simulated to show the method's ability to track multiple deforming interfaces. The bubbles are initially arranged in two layers of four and five bubbles. The domain size is $[10\times20\times10]d_B$, where the gravity is pointing in the $y$-direction. The physical parameters are equivalent to case 3D-I.\\

The bubble shape as well as the velocity field at $t^*=1$ are shown in~\figref{fig:vel_3D9}. To highlight the relative motion of the bubbles, the initial bubble configuration is transposed by the mean translation distance of all bubbles $v_{mean}\cdot t^*$. In the bottom layer, the center bubble is rising at a slower rate compared to its direct neighbors. This is due to the downwash generated by the surrounding bubbles and the confinement. At the same time, the four outer bubbles are pushed outward due to the lateral displacement of fluid by the center bubble. In the top layer, the streamlines around the bubbles are much more symmetric, since the lateral translation is much less pronounced. Additionally, the top layer rises faster due to the downwash it creates for the lower layer.
\section{Summary}
\label{sec:summary}
A coupled level-set lattice Boltzmann two-fluid method on adaptive Cartesian grids for the simulation of liquid-gas multiphase flows was introduced. The new approach effectively combines the strengths of lattice Boltzmann methods and level-set techniques, allowing for accurate tracking of interfaces while addressing the computational challenges caused by complex interfacial dynamics.

The proposed method facilitates the independent selection of the most suitable solution algorithms for each fluid phase, thereby improving accuracy and computational efficiency of the simulations. Different solution algorithms, such as the BGK and cumulant collision kernels can be employed without altering the underlying discrete Boltzmann equation. This is in contrast to virtually all lattice Boltzmann based one-fluid approaches, which typically require explicit modifications of the kinetic equation or collision operators to handle multiphase interactions. We demonstrated the validity of our framework by simulating various test cases, including laminar stratified flows, and single and multiple rising bubbles, which showcased the method's quality to replicate analytical reference solutions with high fidelity.

Moreover, the implementation of adaptive mesh refinement (AMR) allows for optimal resource allocation by dynamically refining computational grids in regions where steep gradients or intricate flow features occur. This capability significantly improves simulation accuracy without causing unnecessarily high computational cost in less critical areas.

Overall, our findings indicate that the coupled level-set lattice Boltzmann method is an efficient and accurate tool to analyze multiphase flow phenomena. Future work will focus on further refining this methodology by investigating more complex scenarios involving bubble break-up, coalescence, and mass transfer.


\bibliographystyle{elsarticle-num} 
\bibliography{lit.bib}

\end{document}